\documentclass[useAMS,usenatbib]{mn2e}
\usepackage{times,amsmath,amssymb,units,graphicx,aas_macros}
\usepackage{multirow,txfonts,natbib,xspace}
\newcommand\cm{\,\rm cm}
\newcommand\m{\,\rm m}
\newcommand\s{\,\rm s}
\newcommand\ms{\,\rm m\,s^{-1}}
\newcommand\g{\,\rm g}

\newcommand\yr{\,\rm yr}
\newcommand\Myr{\,\rm Myr}

\newcommand\km{\,\rm km}
\newcommand\kms{\,\rm km\,s^{-1}}

\newcommand\au{\,\rm au}

\newcommand\perccm{\,\rm cm^{-3}}
\newcommand\Msun{\,\rm M_\odot}

\newcommand\tms{\!\times\!}
\newcommand\cdt{\!\cdot\!}

\newcommand\xx{\hat{{\mathbf x}}}
\newcommand\yy{\hat{{\mathbf y}}}
\newcommand\zz{\hat{{\mathbf z}}}
\newcommand\V{\mathbf v}
\newcommand\B{\mathbf B}
\newcommand\Rp{R_{\rm p}}
\newcommand\vK{v_{\rm K}}
\newcommand\Cs{c_{\rm s}}
\newcommand\rin{r_{\rm in}}
\newcommand\rout{r_{\rm out}}

\newcommand{\rms}[1]{\left<\right.\!#1\!\left.\right>}

\newcommand\Sc{\mathrm{Sc}}
\newcommand\St{\mathrm{St}}
\newcommand{\simgt}%
           {\,\hbox{\lower0.35ex\hbox{$\sim$}\llap{\raise0.35ex\hbox{$>$}}}\,}
\newcommand{\simlt}%
           {\,\hbox{\lower0.35ex\hbox{$\sim$}\llap{\raise0.35ex\hbox{$<$}}}\,}

\newcommand\NIRVANA{\textsc{nirvana}\xspace}
\newcommand\NIII{\textsc{nirvana-iii}\xspace}

\title[Dynamics of planetesimals embedded in turbulent discs]%
      {On the dynamics of planetesimals
        embedded in turbulent protoplanetary discs}
      \author[Nelson \& Gressel]%
             {Richard~P.~Nelson\thanks{E-mail:r.p.nelson@qmul.ac.uk} and %
              Oliver~Gressel\thanks{E-mail:o.gressel@qmul.ac.uk} \\
              Astronomy Unit, Queen Mary, University of London, Mile End Road,
              London E1 4NS, United Kingdom\\ }

\begin{document}

\date{Accepted 1988 December 15. %
      Received 1988 December 14; %
      in original form 1988 October 11}

\pagerange{\pageref{firstpage}--\pageref{lastpage}} \pubyear{2002}

\maketitle

\label{firstpage}

%

\begin{abstract}
Angular momentum transport and accretion in protoplanetary discs are
generally believed to be driven by MHD turbulence {\it via} the
magneto-rotational instability (MRI). The dynamics of solid bodies
embedded in such discs (dust grains, boulders, planetesimals and
planets) may be strongly affected by the turbulence, such that the
formation pathways for planetary systems are determined in part by the
strength and spatial distribution of the turbulent flow.

We examine the dynamics of planetesimals, with radii between $1\m$ --
$10\km$, embedded in turbulent protoplanetary discs, using three
dimensional MHD simulations. The planetesimals experience gas drag and
stochastic gravitational forces due to the turbulent disc. We use, and
compare the results from, local shearing box simulations and global
models in this study.

The main aims of this work are to examine: the growth, and possible
saturation, of the velocity dispersion of embedded planetesimals as a
function of their size and disc parameters; the rate of radial
migration and diffusion of planetesimals; the conditions under which
the results from shearing box and global simulations agree.

We find good agreement between local and global simulations when
shearing boxes of dimension $4H \times 16H \times 2H$ are used ($H$
being the local scale height). The magnitude of the density
fluctuations obtained is sensitive to the box size, due to the
excitation and propagation of spiral density waves. This affects the
stochastic forcing experienced by planetesimals. The correlation time
associated with the stochastic forcing is also found to be a function
of the box size and aspect ratio.

The equilibrium radial velocity dispersion, $\sigma(v_r)$, obtained
depends on the radii, $\Rp$, of the planetesimals. Bodies with
$\Rp=50\m$ achieve the smallest value with $\sigma(v_r) \simeq 20\ms$.
Smaller bodies are tightly coupled to the gas, and boulders with $\Rp
=1\m$ attain a value of $\sigma(v_r)$ similar to the turbulent
velocity of the gas ($\sim 100\ms$).  Equilibrium values of
$\sigma(v_r)$ for bodies larger than $100\m$ are not achieved in our
simulations, but in all models we find rapid growth of the velocity
dispersion for planetesimals of size $1\km$ and $10\km$, such that
$\sigma(v_r) \ge 160\ms$ after a run time of 1200 orbits at a distance
of $5\au$ from the central star. These values are too large to allow
for the runaway growth of planetesimals, and mutual collisions would
lead to catastrophic disruption. Radial migration due to gas drag is
observed for bodies with $\Rp \simeq 1\m$, and is only modestly
affected by the turbulence. Larger bodies undergo a random walk in
their semi-major axes, leading to radial diffusion through the
disc. For our fiducial disc model, we estimate that radial diffusion
across a distance of $\simeq 2.5\au$ would occur for typical
planetesimals in a swarm located at $5 \au$ over a disc life-time of
$5\Myr$. Radial diffusion of this magnitude appears to be inconsistent
with Solar System constraints.

Our models show that fully developed MHD turbulence in protoplanetary
discs would have a destructive effect on embedded planetesimals.
Relatively low levels of turbulence are required for traditional
models of planetesimal accretion to operate, this being consistent
with the existence of a dead zone in protoplanetary discs.

\end{abstract}


%

\section{Introduction}

The dynamical and collisional evolution of planetesimals is a
fundamental issue which needs to be understood if progress is to be
made in developing a theory of planetary system formation.  According
to the core accretion theory, a process which begins with the
collision and sticking of small dust grains within a protoplanetary
disc leads eventually to the formation of kilometre sized
planetesimals \citep{1993Icar..106..190W}.  Although alternative
scenarios have been put forward for planetesimal formation
\citep{1973ApJ...183.1051G, 2007Natur.448.1022J}, which avoid the
requirement for such large bodies to grow {\it via} simple two-body
agglomeration processes, the further growth of planetesimals into
planetary embryos and cores generally requires planetesimals
themselves to accrete {\it via} mutual collisions through a process of
runaway growth, followed by oligarchic growth
\citep{1993Icar..106..210I, 1998Icar..131..171K}.

Rapid runaway growth requires that the velocity dispersion of the
planetesimal swarm remains significantly smaller than the escape
velocity from the surfaces of the largest accreting planetesimals,
ensuring that gravitational focusing is important. For bodies of
radius $10\km$, and with internal densities $\varrho_p \simeq
2\g\cm^{-3}$, the escape velocity is $10\ms$, and scales linearly with
radius. Clearly this is a stringent requirement, which is easily met
within a self--stirring planetesimal swarm whose size distribution is
reasonably shallow, but which may be difficult to satisfy in the
presence of an external source of stirring. One such source may be
turbulence within the protoplanetary disc. Planetary growth times
which rely on mutual collisions between planetesimals occurring at
rates which are determined by the geometric cross section are
prohibitively long, leading to estimated planetary growth times which
are much in excess of typical protostellar disc life times
\citep{2001ApJ...553L.153H}.

A further constraint during the runaway growth phase is that
collisional velocities should be small enough to avoid catastrophic
disruption of planetesimals. For bodies in the 1 - $10\km$ size range,
for which self--gravity starts to become more important than material
strength in holding planetesimals together, collisions between similar
sized bodies with impact speeds which are modestly in excess of the
escape velocity will lead to break up of the planetesimals rather that
accretion and growth. Indeed \citet{1999Icar..142....5B} suggest that
mutual collisions between $1\km$ sized bodies will result in
catastrophic disruption if the impact speeds exceed $\sim 20\ms$,
depending on the material composition of the impactors. In a more
recent study, \citet{2009ApJ...691L.133S} suggest reduced impact
speeds of $\sim 10\ms$ will be destructive. Once again, we see that an
external source of planetesimal stirring may prevent the rapid growth
of planetary mass bodies by the accretion of planetesimals.

The canonical mass accretion rate for T Tauri stars is $\sim
10^{-8}\Msun\yr^{-1}$ \citep{2004AJ....128..805S}. Such accretion
rates require a source of anomalous disc viscosity and angular
momentum transport to operate, generally thought to be turbulence. The
most likely source of disc turbulence is the magneto-rotational
instability \citep[MRI,][]{1991ApJ...376..214B} which has been shown to
develop into full non linear MHD turbulence in numerous studies, using
both local shearing box simulations \citep{1995ApJ...440..742H}, and
global simulations \citep{1998ApJ...501L.189A,
2001ApJ...554..534H,2003MNRAS.339..983P}. The nature and saturation
state of MHD turbulence generated by the MRI is the subject of
on-going study \citep{2007A&A...476.1113F, 2007A&A...476.1123F}. In
this work, we use the dependence of the turbulent stresses and density
fluctuation amplitude on the strength of the net component of the
magnetic field to examine the evolution of planetesimals in discs with
different levels of turbulence. We use simple disc models, which
neglect non-ideal MHD effects and vertical stratification. As such,
this is the first in a series of papers in which we examine how
turbulence affects the dynamics of planetesimals embedded in turbulent
discs. Future papers will explore the effects of vertical
stratification and dead zones.

There have been numerous studies of planets embedded in turbulent
protoplanetary discs. \citet{2003MNRAS.339..993N} and
\citet{2003ApJ...589..543W} examined the formation of gaps by Jovian
mass planets, and the migration torques exerted by the disc on the
planet. \cite{2004MNRAS.350..849N} performed global simulations of low
mass planets embedded in turbulent discs. They showed that such bodies
are subject to fluctuating torques which should induce stochastic
migration, and suggested that this might provide a means of mitigating
against the rapid type I migration expected to occur for low mass
planets \citep{1997Icar..126..261W}.  \citet{2004ApJ...608..489L}
published a similar study using analytical fits to MHD simulations,
and reached similar conclusions. \citet{2004MNRAS.350..829P} presented
results from both global and local shearing box simulations containing
both high and low mass planets, and showed good agreement between the
simulation set-ups for predicting the transition between linear and
non linear disc response to the presence of an embedded planet. In a
follow-up paper, \cite{2005A&A...443.1067N} examined the orbital
evolution of low mass embedded planets, showing that over simulations
run times of $\simeq 100$ orbits, turbulence induces stochastic
migration for planets in the range 1 -- 10 M$_{\oplus}$, and induces
the growth of orbital eccentricity. In more recent work,
\citet{2007ApJ...670..805O} examined the stochastic forces experienced
by planets in stratified disc models with and without dead zones using
shearing box simulations, and \citet{2009ApJ...707.1233Y} examined the
orbital evolution of swarms of test particles embedded in non
stratified turbulent discs.

A significant volume of related work has examined the influence of
disc turbulence on embedded planets and planetesimals using
prescriptions or simple models for the effects of
turbulence. \citet{2006ApJ...647.1413J} developed a Fokker-Planck
description for the stochastic evolution of planets, and examined the
survival probabilities of distributions of planets subject to type I
migration and superposed stochastic migration. A similar study has
been published recently by \citet{2009ApJ...701.1381A}.
\citet{2007Icar..188..522O} have used N-body simulations plus a
prescription for stochastic forcing to examine the effects of
turbulence on terrestrial planet formation.
\citet{2008ApJ...686.1292I} used a similar prescription of turbulent
forcing and examined the growth of eccentricity for planetesimals,
exploring in particular the possibility of reaching catastrophic
disruption velocities.  \citet{2008ApJ...683.1117A} and
\citet{2009A&A...497..595R} examined the stability of mean motion
resonances for pairs of planets embedded in turbulent discs, and
\citet{2010ApJ...709..759B} have examined the saturation of corotation
torques in turbulent discs by means of hydrodynamic simulations
subject to turbulent stirring.

In this paper, we examine in detail the orbital evolution of
planetesimals of different size (ranging between $1\m$ and $10\km$)
embedded in turbulent disc models by means of three dimensional MHD
simulations. A key issue that we explore is the set of conditions and
numerical parameters that provide good agreement between local
shearing box simulations and global simulations.  We find that it is
possible to obtain good agreement between these two numerical set-ups,
provided that the shearing box dimensions are chosen
appropriately. Other important issues that we examine include the
growth of the velocity dispersion of embedded swarm of planetesimals,
and the saturation value of this velocity dispersion as a function of
planetesimal size due to a balance being achieved between gas drag and
turbulent forcing. We explore the implications of our results for the
efficiency of runaway growth of planetary embryos, and the possibility
that planetesimals may enter a phase of catastrophic disruption
through mutual collisions.  We also examine the rate at which
planetesimals migrate due to both gas drag induced radial drift
\citep{1977MNRAS.180...57W}, and due to diffusion caused by the
fluctuating gravitational field of the turbulent disc. We examine
under which conditions each of these processes are dominant, and we
explore the implications of our results for the radial drift of
planetesimals in the Solar nebula and limits that might be placed on
the magnitude of turbulent fluctuations which were present during the
early phases of Solar System formation.

This paper is organised as follows. In Sect.\ref{Sec:Models} we
describe the numerical set-up and the parameters of the disc
models. In Sect.~\ref{sec:Results} we present our results. In
Sect.~\ref{sec:Conclusions} we discuss our work in the context of
previous work and draw our conclusions.

%

\section{Model description}
\label{Sec:Models}

We perform self-consistent simulations of hydromagnetic turbulence
using two different set-ups: shearing box simulations which represent
a local patch of a protoplanetary disc; global disc models which
simulate a larger section of a protoplanetary disc and include the
full set of curvature terms in the equations of motion.  A major goal
of this work is to compare the results of these two different
approaches.

A key question that needs to be addressed is for what dimensions of
the the shearing box, in units of the local scale height $H$, do
density fluctuations created by the turbulence reach a converged
amplitude and spectrum, and do these match the results of global
models. To make the comparison as straightforward as possible, we
neglect vertical stratification and assume an isothermal equation of
state.

In both configurations, the hydromagnetic turbulence is driven via the
non-linear development of the magneto-rotational instability. At
present, the issue of the saturated amplitude of MRI turbulence
remains unresolved and is a topic of active research
\citep{2007A&A...476.1113F, 2007A&A...476.1123F}.  In the absence of a
better alternative, we therefore adopt a practical perspective and
impose a net vertical or azimuthal flux, for which numerical
convergence can be obtained \citep{2009arXiv0909.1570D}.  Neglecting
the dependence on the magnetic Prandtl number
\citep{2007MNRAS.378.1471L}, we furthermore restrict ourselves to the
case of ideal MHD.  This approach is justified by the observed
correlation between the strength of the turbulence and the amplitude
of the resulting density fluctuations
\citep{2009ApJ...707.1233Y}. This means that we regard the strength of
the imposed field as a control parameter which can be tuned to vary
the turbulence amplitude in the local and global context. The global
cylindrical disc models are computed with a modified version of the
original finite difference code \NIRVANA
\citep{1997CoPhC.101...54Z}. For the local shearing box models, we
make use of the newly developed second-order Godunov code \NIII
\citep{2004JCoPh.196..393Z,2008CoPhC.179..227Z}.


\subsection{Numerical methods -- local model}
\label{sec:local-model}

For our standard model, we adopt a box size\footnote{See
Sect.~\ref{sec:torques} for a discussion on the effect of the box
size.} of $4\tms16\tms2$ pressure scale heights $H$ at a resolution of
$32$ grid points per $H$. Boundary conditions are periodic in the
azimuthal ($y$) and vertical ($z$), and sheared-periodic in the radial
($x$) direction.  The initial net vertical magnetic field corresponds
to a plasma parameter $\beta \simeq 6000$, resulting in a typical
saturation level $\alpha \simeq 0.05$ of the turbulence, where
$\alpha$ is the effective viscosity parameter
\citep{1973A&A....24..337S}, and $\beta$ is ratio of thermal to
magnetic pressure.

Because the gas drag forces acting on massive particles depend on the
actual physical value of the gas density, we have to prescribe a set
of conversion factors to link our model to a representative
protoplanetary disc. We chose a fiducial radius $R_0=5\au$, and a
geometric disc thickness of $H/R=0.05$ at $R=1\au$. Note that this
aspect ratio is scaled with $R^{1/4}$ to be consistent with the
Hayashi minimum mass solar nebula
\citep[MMSN,][]{1981PThPS..70...35H}, yielding a value of $\simeq
0.075$ at $5\au$. Furthermore, we chose a slightly higher average mass
density than in this model to yield a column density $\Sigma=160\,{\rm
g\,cm^{-2}}$ and sound speed $\Cs=1 \kms$ comparable to the global
simulations.

For the local model, we evolve the following set of non-linear partial
differential equations
\begin{eqnarray}
      \partial_t\varrho +\nabla\cdt(\varrho \V) & = & 0\,, \nonumber\\
      \partial_t(\varrho\V) +\nabla\cdt
          \left[ \varrho\mathbf{vv} + (p + \frac{\B^2}{2\mu})\,I 
            -\frac{\mathbf{BB}}{\mu} \right] & = & \!\!
            -2\varrho\Omega\,\zz\times\left(\V+q\Omega x\,\yy \right)\nonumber\\
      \partial_t \B -\nabla\tms(\V\tms\B) & = & 0\,,
\label{eqn:motion}
\end{eqnarray}
comprising the standard formulation of ideal MHD in the shearing box
approximation, and where we have assumed an isothermal equation of
state $p=\varrho\,\Cs^2$ and neglected the effects of
stratification. The two momentum source terms are the Coriolis force
$-2\Omega\zz\tms\V$ in the locally corotating frame, and the tidal
term $2 q\Omega^2 x\,\xx$, with shear-parameter $q=3/2$, describing
the linearised effect of the Keplerian rotation.\footnote{The variable
$x=R-R_0$ is the radial displacement from the box centre.} Care has
been taken in implementing the source terms to minimise the error in
the epicyclic mode energy \citep{2007CoPhC.176..652G}, albeit not to
the extent where it is conserved to machine accuracy
\citep{2010ApJS..189..142S}.


\subsubsection{Numerical scheme \& orbital advection}

As has been recently demonstrated by \cite{2010arXiv1003.0018B}, the
adequate modelling of the MRI with finite volume codes depends on the
reconstruction strategy used and, in particular, on the ability of the
Riemann solver to capture the Alfv{\'e}n mode. To improve the
treatment of discontinuities in the Godunov scheme, we therefore
extended \NIII with the Harten--Lax--van Leer Discontinuities (HLLD)
approximate Riemann solver proposed by \cite{2005JCoPh.208..315M}.

In accordance with its finite volume approach, the \NIII code
evaluates the components of the electromotive force (EMF) at cell
interfaces. Since the discretisation of the constrained transport
algorithm intrinsically requires edge-aligned EMFs, some sort of
interpolation is required. In its original form, \NIII implements the
arithmetic average proposed by \cite{1999JCoPh.149..270B}. As
discussed in Sect. 3.2 of \cite{2005JCoPh.205..509G}, this approach
however lacks the required directional biasing to guarantee the
stability of the numerical scheme. It has further been demonstrated by
\cite{2009arXiv0906.5516F} that this can lead to artificial growth of
instabilities in the context of net-flux MRI, and we have reproduced
this result. To resolve this issue, we have successfully implemented
and tested the upwind reconstruction procedure of
\cite{2005JCoPh.205..509G}.

Following the long-term evolution of a shearing flow in boxes of
substantial radial extent puts high demands on computational
resources. For a Keplerian rotation profile, the background flow
becomes super-sonic for $L_x > 4/3 H$. This implies that for
increasingly larger boxes the numerical time-step, defined by the
Courant condition, becomes dominated by the unperturbed shear
profile. To circumvent this undesirable constraint, it becomes
mandatory to split-off the transport term due to the background
profile.

The shear transport is usually implemented in terms of an
interpolation step. This was first introduced in cylindrical geometry
and termed FARGO by \cite{2000A&AS..141..165M}. Later, the method was
adopted to the local shearing box model by \cite{2001ApJ...553..174G}.

A rather intricate extension for the induction equation that requires
mapping of the magnetic field components has been proposed by
\cite{2008ApJS..177..373J}. We here follow the (much simpler)
constrained transport approach proposed by \cite{2010ApJS..189..142S},
which by construction preserves the solenoidal constraint.

For our implementation of the orbital advection scheme, we
operator-split the advection step from the Runge-Kutta time
integration of the remaining terms. For the interpolation of the fluid
variables, we make use of the high-order Fourier scheme (SAFI)
described in appendix~B of \cite{2009ApJ...697.1269J}.

We similarly apply SAFI to obtain the non-integer part of the
line-integrals which contribute the circulation of the electric fields
entering the induction equation \citep[cf. Eqs.~(61) and (62)
in][]{2010ApJS..189..142S}. The treatment of the magnetic source term
in the total energy equation can be successfully avoided if the
magnetic energy is removed from the total energy during the
interpolation. The implementation has been tested with the advection
of a field loop \citep{2008JCoPh.227.4123G}, and the exact wave
solution given in \citet{2006ApJ...652.1020B}.

Using SAFI rather than slope-limited linear interpolation, efficiently
reduces the dissipation due to the transport step, and moreover its
dependence on position \citep[see][]{2009ApJ...697.1269J}. In fact the
scheme adds so little dissipation that the TVD requirement might be
violated. To formally make the interpolation total variation
diminishing, we therefore discard the Fourier mode corresponding to
the Nyquist frequency.


\subsubsection{Particle dynamics}

In this paper, we restrict ourselves to the study of how disc
turbulence affects embedded particle populations. Neglecting their
back reaction on the flow, particles are hence treated as passive test
bodies which do not interact mutually either through physical
collisions or gravity. Under these assumptions, we ignore the
possibility of increasing the velocity dispersion of particles via
mutual gravitational scattering. While this effect might become
important for $\sim 10^2 \km$-sized objects, it simply adds to the
external stirring. Physical collisions between planetesimals, however,
can provide a source of damping. This effect was considered by
\cite{2008ApJ...686.1292I}, and was found to be important in
determining the equilibrium velocity dispersion only for bodies with
size $< 1 \km$, and so we do not consider this effect in this paper.
Moreover, because the particles cannot exert drag forces on the gas,
our approach excludes collective effects such as the streaming
instability \citep{2005ApJ...620..459Y}, which is a focus of current
numerical studies \citep{2007ApJ...662..613Y,%
2009MNRAS.397...24B,2010arXiv1005.4982B,2010JCoPh.229.3916M}.

We include different species of particles to quantify various aspects
of the flow. Firstly, massless tracer particles (which instantaneously
follow the gas velocity) measure the Lagrangian diffusion of the
flow. This is relevant for small dust grains which are tightly coupled
to the gas. Secondly, we include particles representing
planetesimals. These particles interact with the flow {\it via} the
gravitational potential produced by the gas density, and through the
aerodynamic drag force.  The relative importance of these effects is
expected to change for planetesimals in the size range $1\m$ to
$10\km$, which are the subject of this study.
Finally, for the purpose of comparison, and as a proxy for larger
objects (e.g. small protoplanets), we include swarms of particles
which experience gas gravity but are not subject to an aerodynamic
drag force. With the exception of the tracers, all particles are
subject to the local dynamics, i.e., they experience the Coriolis
force in the rotating frame and the tidal force stemming from the
local expansion of the Keplerian rotation profile. As a consequence,
planetesimals generally perform epicyclic oscillations of fluctuating
amplitude, around a stochastically migrating guiding centre.

\subsection{Numerical method - global model}
In the global simulations we solve essentially the same set of
equations for ideal MHD as described in Sect.~\ref{sec:local-model}
for the shearing box runs, except that we adopt cylindrical
coordinates ($r$, $\phi$, $z$) (see \citet{2005A&A...443.1067N} for a
full description).  The simulations are performed in a rotating frame
with angular frequency equal to the Keplerian frequency at the
midpoint of the radial computational domain.  We use a locally
isothermal equation of state
\begin{equation}
P(r)= \Cs(r)^2\ \varrho,
\end{equation}
where $\Cs(r)$ denotes the sound speed which is specified as a fixed
function of $r.$ The models investigated may be described as
cylindrical discs \citep[e.g.][]{2001ApJ...554..534H}, in which the
gravitational potential is taken to depend on $r$ alone. Thus the
cylindrical disc models do not include a full treatment of the disc
vertical structure. Models of this type are employed due to the high
computational overhead that would be required to resolve fully the
disc vertical structure of a stratified model.

The global simulations presented in this paper were performed using an
older version of \NIRVANA, which uses an algorithm very similar to the
\textsc{zeus} code to solve the equations of ideal MHD
\citep{1997CoPhC.101...54Z, 1992ApJS...80..753S}. This scheme uses
operator splitting, dividing the governing equations into source terms
and transport terms. Advection is performed using the second-order
monotonic transport scheme \citep{1977JCoPh..23..276V}, and the
magnetic field is evolved using the Method of Characteristics
Constrained Transport \citep{1995CoPhC..89..127H}.

 \subsubsection{Gas disc model}
 \label{gasdisc}
The main aim of this paper is to examine the orbital evolution of
planetesimals and smaller bodies (boulders) in turbulent discs, where
the disc turbulence has achieved a well-defined steady state.  Another
aim is to compare local shearing box runs with global simulations. To
achieve these aims, most of the global disc models were chosen to have
a relatively narrow radial extent, and azimuthal domains running
between $0 \le \phi \le \pi/2$. The turbulent stresses generated
within a global disc lead to significant changes in the radial density
distribution, such that the properties of the disc deviate
substantially from a statistical steady state over runs times of 100s
of orbits \citep{2003MNRAS.339..983P}.  In order to overcome this, we
have introduced an additional equation to be solved alongside
Eqs.~(\ref{eqn:motion}) in the global simulations, whose purpose is
to maintain a roughly constant surface density profile during the
runs:
\begin{equation}
  \frac{d \varrho(t)}{dt} = -\frac{\varrho(t)-\varrho_0}{\tau_{\rm m}}.
\label{massadd}
\end{equation}
Here $\varrho(t)$ is the density at each spatial location in the disc
at time $t$, $\varrho_0$ is the initial density defined at each
location in the disc, and $\tau_{\rm m}$ is the characteristic time on
which the local perturbed density evolves back toward its original
value. Clearly if $\tau_{\rm m}$ is shorter then any relevant local
dynamical time in the disc, then perturbations will be damped very
quickly and the disc will not be able to achieve a turbulent state.
Similarly, if $\tau_{\rm m}$ is much longer than the longest (viscous)
evolution time in the disc, the global density profile will be able to
evolve to be very different from the initial one. By choosing
$\tau_{\rm m}$ to be intermediate between these two extremes, we find
that our global disc models are able to develop a well defined
turbulent state, in which the turbulence is able to maintain an
approximate statistical steady state over long time scales ($> 10^3$
orbits). We have found that $\tau_{\rm m}=50$ orbits measured at the
radial midpoint of the disc model provides a model with the desired
properties.

A possible alternative to solving Eq.~(\ref{massadd}) would be to feed
in mass at the outer radial boundary at the requisite rate.  Such an
approach works well as a means of generating a steady disc when the
disc is laminar and employs the $\alpha$-model for viscosity
\citep{2002A&A...387..605M}. In a turbulent disc, where there is
substantial temporal and spatial variation in the viscous stress,
however, such an approach may not be effective at maintaining a
well-defined surface density profile.

\subsubsection{Initial and boundary conditions}
Each global model has a value for the inner and outer radii of the
computational domain, $\rin$ and $\rout$, a value for the height of
the upper and lower ``surfaces" of the disc, $Z_{\rm min}$ and $Z_{\rm
max}$, and minimum and maximum values of the azimuthal angle,
$\phi_{\rm min}$ and $\phi_{\rm max}$. These values are tabulated in
table~\ref{table1}, along with some of the physical parameters
described below.  The resolution used in each simulation covering
$\pi/2$ in azimuth was ($N_r$, $N_{\phi}$, $N_z$) = (160, 320, 40).
For the model covering $2 \pi$ in azimuth $N_{\phi}=1280$.
This corresponds to 10 cells per mean scale height
in the radial and azimuthal directions, and 20 cells per scale
height in the vertical domain. When considering the behaviour of 
the correlation time for the fluctuating torques in 
Sect.~\ref{sec:trq_acf}, we also ran models with double and 
quadruple the resolution in the radial and azimuthal directions.

The unit of length in our simulations is taken to be
$2\au$, such that a radial distance of $r=2.5$ in computational units
corresponds to $5\au$. The unit of mass is assumed to be the Solar
mass. When we discuss the temporal evolution of our models later in
the paper, we adopt a time unit which is equal to the orbital period
at $r=2.5$ (equivalent to $5\au$), this being the midpoint of
our radial domain.

The disc models we adopt are similar locally to the minimum mass solar
nebula model \citep[MMSN,][]{1981PThPS..70...35H} at a distance of
$5\au$ from the star. All models have a constant aspect ratio $H/r$,
and all but one model has $H/r=0.05$. Model G5 has $H/r=0.075$ for the
purpose of comparing directly with the local shearing box models
(where this value of $H/r$ was used). Initially the density is
constant, such that the surface density $\Sigma=150\g\cm^{-3}$.

The initial magnetic field in the global runs is purely toroidal, with
the local field strength being determined from the local plasma
$\beta$ parameter ($\beta=P_{\rm gas}/P_{\rm mag}$). The value of
$\beta$ used in each run is tabulated in table~\ref{table1}. The field
is introduced at all locations in the disc, except near the radial
boundaries where the field is set to zero for $r-\rin<0.1$ and
$\rout-r<0.1$. The initial disc velocity is determined according to
\begin{equation}
  v_{\phi}= r\ \sqrt{\frac{G M_*}{r^3} \left[ 1-
      \left(\frac{H}{r}\right)^2\right]}\ ,
\end{equation}
where $M_\star$ is the mass of the central star, with the other
velocity components being set to zero. Prior to a run being initiated,
however, all velocity components are seeded with random noise with an
amplitude equal to 5 \% of the local sound speed.

Each disc model described in table~\ref{table1} was evolved until the
turbulence reached a quasi-steady state before the planetesimals were
inserted. Most runs included planetesimals of size $10\m$, $100\m$,
$1\km$ and $10\km$, with each size being represented by 25
particles. The planetesimals were distributed randomly in a narrow
annulus centred on the computational radius $r=2.5$ (equivalent to
$5\au$ in physical units) of width $\Delta r=0.2$, with their initial
velocities calculated such that they are on circular orbits under the
influence of the instantaneous gravitational potential of the central
star and turbulent disc.

We adopt periodic boundary conditions at the vertical and azimuthal
boundaries, and reflecting boundary conditions at the radial
boundaries. Furthermore, we use wave damping boundary conditions in
the vicinity of the radial boundaries using the scheme described in
\cite{2006MNRAS.370..529D}. This scheme relaxes the velocity and
density near the boundaries toward their initial values on a time
scale equal to 10 \% of the local orbital period.


\subsection{Gravitational forces}
\label{sec:f_grav}

Both the local and the global models calculate the gravitational
potential of the gas at the position of the particles via direct
summation. This is computationally favourable as long as one is only
interested in a relatively small number of particles. Since we do not
consider collective effects, an ensemble of a few tens of members is
usually sufficient to reasonably determine the time-averaged
distribution of their positions and velocities.

For the integration of the gravitational force, the mass contained
within a grid cell is treated as a point source located at the cell
centre. To avoid artifacts due to close encounters, we apply a common
smoothing length formalism with a parameter $b=(\delta_x^2+\delta_
y^2)^{\nicefrac{1}{2}}$ equal to the diagonal across the cell.

As can be seen in Fig.~2 of \cite{2009MNRAS.397...64H}, the turbulence
driven by long-wavelength MRI modes leads to the formation of strong
density waves, which develop very little structure along the vertical
direction. Accordingly, we neglect the vertical component of the
forces and adopt a cylindrical description where the gravity now only
depends on the vertically integrated column density. This approach
greatly reduces the computational demand and is consistent with
neglecting the vertical density stratification.  As we will see in
Sect.~\ref{sec:torques}, this 2D treatment enhances the gravity forces
by a factor of roughly two.

\begin{figure}
  \center\includegraphics[width=0.8\columnwidth]{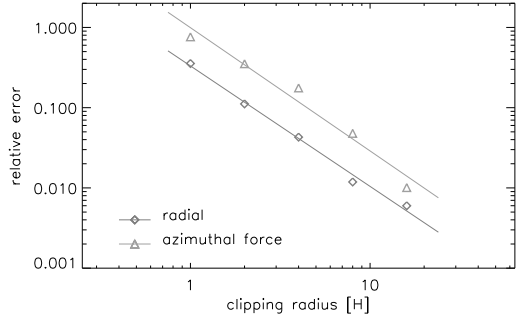}
  \caption{Convergence study regarding the computation of the
    gravitational force via direct summation; the convergence order is
    approximately $3/2$.}
  \label{fig:fg_conv}
\end{figure}

When considering local shearing box simulations in particular, a
question that needs to be answered is how large a shearing box is
required before the gravitational forces from a turbulent disc are
converged.  In Fig.~\ref{fig:fg_conv}, we plot the relative error of
the gravity force when integrating over spheres with increasing
radius. We see that, for values accurate at the percent level, one
requires a sphere of influence with a radius of about ten pressure
scale heights. As expected, forces in the azimuthal direction are
affected more strongly by long-range contributions. Because the net
effect is determined by density fluctuations, the convergence is
weaker than the $r^{-2}$ dependence for Newtonian gravity. We
consequently find a smaller convergence order of about $3/2$. In our
local simulations, we adopt a sphere (cylinder) with $r_{\rm cl}=16H$,
within which we compute the gravitational acceleration on a planetesimal,
and suitably extend the domain by mirroring ghost domains according to
the sheared periodicity.

In the global simulations the gravitational field is computed by
summing over all grid cells. Most simulations were run using an
azimuthal domain of $\pi/2$, although the planetesimal orbits cover
the full $2\pi$ domain. When calculating the gravitational field
experienced by the planetesimals, additional copies of the disc, each
shifted by $n\pi/2$, where $n \in \{1,2,3\}$, are used to mimic a disc
which covers the full $2\pi$ in azimuth.  One run with a full $2 \pi$
azimuthal domain was run to check that the above procedure gives
accurate results.

\subsection{Gas drag}
\label{sec:gas-drag}

For the gas drag, we use the usual formulae for the Stokes and Epstein
regimes \citep{1977MNRAS.180...57W, 2004AJ....128.1348R}.  In the
Epstein regime, the drag force is given by
\begin{equation}
  {\bf F}_{\rm drag} = ({\bf v}_{\rm g} - {\bf v}_{\rm p})\ \tau_{\rm
  s}^{-1}\,,
  \label{eqn:epstein}
\end{equation}
with the stopping time
\begin{equation}
  \tau_{\rm s} = \frac{\varrho_{\rm p} \Rp}{\varrho \Cs}\,,
\end{equation} 
where $\Rp$ is the physical radius of the planetesimal, $\varrho$ is
the gas density at the position of the planetesimal, $\varrho_{\rm p}$
is the internal density of the planetesimal, ${\bf v}_{\rm p}$ is the
planetesimal velocity, ${\bf v}_{\rm g}$ is the gas velocity. In the
Stokes regime it may be written
\begin{equation}
  {\bf F}_{\rm drag}= \frac{1}{2}C_{\rm D} \pi \Rp^2 \varrho\ |{\bf
  v}_{\rm p} - {\bf v}_{\rm g}|\ ({\bf v}_{\rm p} - {\bf v}_{\rm g})
\label{eqn:stokes}
\end{equation}
where $C_D$ is the drag coefficient, which takes the values
\begin{equation}
C_D = 
\begin{cases}
  \ 24.\ {\cal R}_e^{-1} & \text{${\cal R}_e < 1$} \\
  \ 24.\ {\cal R}_e^{-0.6} & \text{$1 < {\cal R}_e \le 800$} \\
  \ 0.44 & \text{${\cal R}_e > 800$}
\end{cases}
\label{eqn:C_D}
\end{equation}
where ${\cal R}_e$ is the Reynolds number of the flow around the
planetesimals, defined by
\begin{equation}
  {\cal R}_e = 2 \Rp v_{\rm pg}/\nu_{\rm p}
\end{equation}
where $v_{\rm pg}=|{\bf v}_{\rm p} - {\bf v}_{\rm g} |$, and the
molecular viscosity of the flow around the planetesimal is given by
\begin{equation}
  \nu_{\rm p}=\lambda \Cs/3
  \label{eqn:nu-global}
\end{equation}
in most of the global simulations we have performed. It should be
noted that in the local simulations, however, the molecular viscosity
was defined by
\begin{equation}
  \nu_{\rm p}=\lambda \Cs/2,
  \label{eqn:nu-local}
\end{equation}
so we have run one global model (G5) adopting this value.  The
molecular mean free path $\lambda = (n \sigma_{{\rm H}_2} )^{-1}$,
where $n=\varrho/(\mu m_H)$ is the number density of particles, $\mu$
is the mean molecular weight, and $m_H$ is the mass of the hydrogen
atom. All but one global simulation adopted the assumption that the
disc gas is composed entirely of molecular hydrogen, giving $\mu=2$.
The local shearing box models adopted a value of $\mu=2.4$, so we have
run one global model (G5) using this value in order to provide a
direct comparison between local and global models.  We adopt a value
of $\sigma_{{\rm H}_2} =10^{-15}\cm^2$ for the collision cross section
of molecular hydrogen \citep{2004AJ....128.1348R}.  Trilinear
interpolation is used to obtain the gas density and velocity at the
position of each planetesimal.

The time integration of the particle motion in the shearing box
simulations was performed as follows.  Because the gas drag relations
are of the form $v^\alpha$ in the relative velocity $v=|v_{\rm
g}-v_{\rm p}|$, the update can be performed analytically. In the case
$\alpha=1$, the relevant time scale $\omega^{-1}$ is just the
classical stopping time, and the decay is exponential. The update from
time $t_n$ to time $t_{n+1}$ with timestep $\delta t$ can then be
written as
\begin{equation}
  v_{n+1} = v_n\,e^{-\omega\,\delta t}\,.
\end{equation}
For $\alpha\ne 1$, we obtain power-law solutions for the damping of
the relative velocity, which can be expressed as
\begin{equation}
  v_{n+1} = \left[ v_n^{1-\alpha} \left( 1 +
            (\alpha-1)\,\tilde{\omega}\,\delta t\right)
            \right]^{\frac{1}{1-\alpha}}\,,
\end{equation}
where $\tilde{\omega}^{-1}$ is now a generalised ``stopping time''
defined by $\tilde{\omega}=f_{\rm dr}\,v^{-1}$, with $f_{\rm dr}$ the
specific drag force acting on the particle.

In the global simulations, the particles were evolved using a
fifth-order Runge-Kutta scheme \citep{1996nuco.book.....P}.

%

\section{Results}
\label{sec:Results}

We organise the discussion of our results by first describing the
evolution of the disc models. We then describe the evolution of the
velocity dispersion (or equivalently eccentricity) of embedded
planetesimals, followed by a discussion of their migration through
changes in the semimajor axes.


\subsection{Hydromagnetic turbulence - local model}
\label{sec:MHD_turb}

\begin{figure}
  \center\includegraphics[width=\columnwidth]{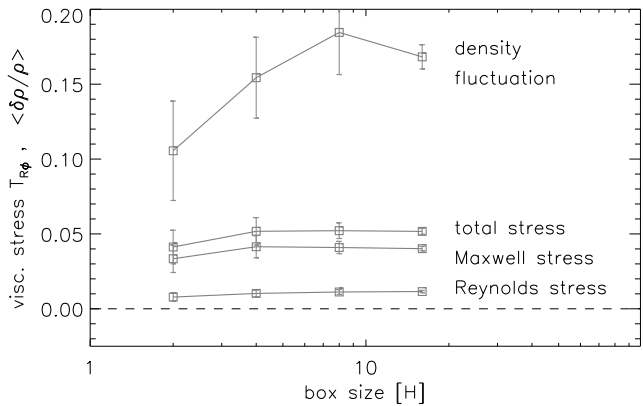}
  \caption{Box size-dependence of key indicators characterising the
    hydromagnetic turbulence: While both the Reynolds and Maxwell
    contributions to the turbulent stress are well converged for
    azimuthal box sizes above $L_y\simgt 4H$, the effected density waves
    are clearly suppressed in too small boxes. Converged values for
    the relative density fluctuation can only be obtained for
    $L_y\simgt 8H$. This is directly reflected in the gravitational
    torques acting on the particles (cf. Fig.~\ref{fig:boxsz_trq}).}
  \label{fig:boxsz_mhd}
\end{figure}

Restricting one's consideration to a local approach, one might
na\"ively think that a relatively small box should suffice to capture
the relevant dynamics \citep[see][for a general discussion on
limitations of the local approximation]{2008A&A...481...21R}. As can
be seen in Fig~\ref{fig:boxsz_mhd}, this notion is supported by
looking at the turbulent stresses created by the saturated MHD
turbulence. Taken as the only criterion, one arrives at the conclusion
that box sizes of $\simgt 4H$ are sufficient to study the local
stirring of particles -- but this is misleading as the gravitational
torques acting on the planetesimals are strongly affected by spiral
density waves, which arise as a secondary feature of the vigorously
driven turbulence in a shearing background.

\begin{figure}
  \center\includegraphics[width=\columnwidth]{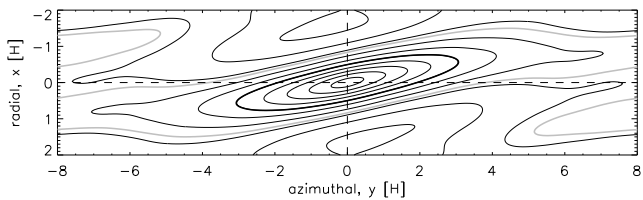}
  \caption{Two-dimensional autocorrelation function of the vertically
    integrated gas density, computed from six uncorrelated snapshots
    of a simulation with box size $4H\tms16H\tms2H$. The thick black
    contour indicates a value of $e^{-1}$ while the thick grey line
    marks the first zero-crossing.}
  \label{fig:rho_acf}
\end{figure}

In a series of two papers, \cite{2009MNRAS.397...52H,%
2009MNRAS.397...64H} study the mechanisms by which spiral density 
waves are excited in a differentially rotating fluid. One important
conclusion of their work is that a minimum azimuthal extent $L_y\simgt
6H$ is required to properly capture the dynamics of spiral waves. This is
exactly what we see when looking at the autocorrelation function (ACF)
of the vertically integrated density, as plotted in
Fig.~\ref{fig:rho_acf}. Here the thick black line indicates the
contour where the correlation falls off to $e^{-1}$, and we see that
density structures are predominantly trailing waves with an azimuthal
extent of about six pressure scale heights. The aspect ratio of the
approximate ellipse is about one sixth, indicating that convergence
in the radial direction should be obtained much earlier (cf. right
panel of Fig.~\ref{fig:boxsz_trq}).

Returning to Fig.~\ref{fig:boxsz_mhd}, we see that this requirement is
reflected in the relative rms density fluctuations $\rms{\delta
\varrho/\varrho}$ measured in the saturated turbulent state (uppermost
curve in Fig.~\ref{fig:boxsz_mhd}). Increasing the box size from $2H$
to $8H$, results in an increase of $\sim 75\%$ in the relative rms
fluctuations. This is considerably larger than the $\sim 25\%$
increase found by \cite{2009ApJ...707.1233Y}, who performed a similar
study. As the authors themselves mention, even this not particularly
dramatic effect seems to strongly affect particle stirring. This can
be understood when taking into account the central finding of the
simulations undertaken in \cite{2009MNRAS.397...64H}, namely that
spiral density waves quickly grow into the non-linear regime where
they develop steep shock-like features. It seems natural that the
resulting intermittent density structure creates highly fluctuating
torques, gravitationally enhancing the turbulent velocity dispersion
of embedded planetesimals.  We examine this issue in more detail in
Sect.~\ref{sec:torques} below.


\subsection{Hydromagnetic turbulence - global models}
\label{global-disc-models}

\begin{figure}
  \center\includegraphics[width=0.9\columnwidth]{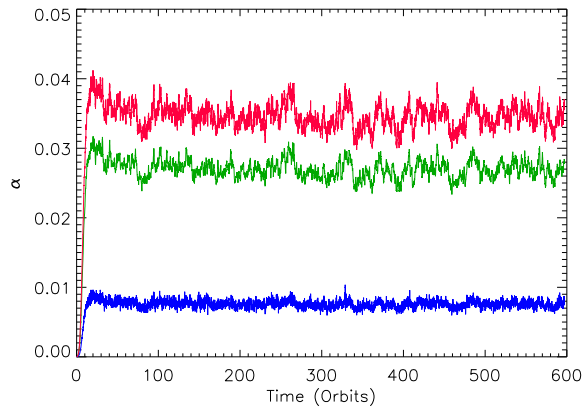}
  \caption{Contributions of the Reynolds (\emph{lower}) and Maxwell
    (\emph{middle line}) stresses to the effective $\alpha$ parameter
    (\emph{upper line}) versus time for model G1.}\label{alphaplot1}
\end{figure}

\begin{figure}
  \center\includegraphics[width=0.9\columnwidth]{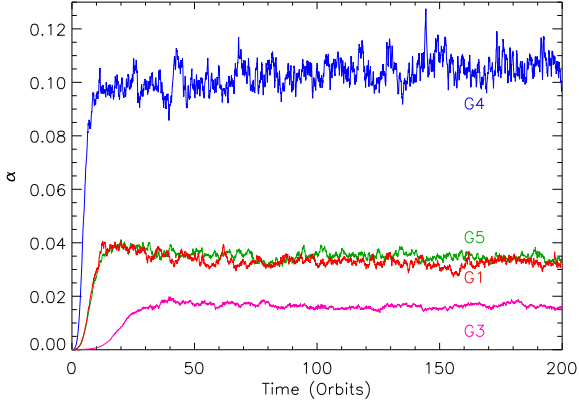}
   \caption{Evolution for the total $\alpha$ value for the models G1,
     G3, G4 and G5 described in table~\ref{table1}.}
   \label{alphaplot2}
\end{figure}

Prior to inserting the planetesimals in the global disc models, we
allow the MRI to develop into quasi-steady non linear MHD turbulence,
with well defined volume averaged stresses operating.  An example of
the time history of the Maxwell, Reynolds and total volume averaged
stress (taken from model G0/G1) is presented in Fig.~\ref{alphaplot1},
and the total stress generated in each of the models as a function of
time is shown in Fig.~\ref{alphaplot2}. It is clear that the models
have evolved to a quasi-steady state. The turbulence generates a
distribution of density and surface density fluctuations which are
well fitted by Gaussian distributions. The computed standard
deviations for these Gaussian fits (which we denote as $\langle \delta
\varrho/ {\varrho} \rangle$ and $\langle \delta \Sigma / { \Sigma}
\rangle$), are tabulated in table~\ref{table1} for each disc model,
along with the time averaged stress parameter $\alpha$. As can be seen
from table~\ref{table1}, our models generate $\alpha$ values in the
range $0.017 \le \alpha \le 0.101$, with corresponding rms values for
$\langle \delta \varrho / { \varrho} \rangle$ in the range $0.101 \le
\langle \delta \varrho / { \varrho} \rangle \le 0.266$.

In terms of physical parameters, the global model which is most
similar to the shearing box simulation is run G5. We see that $\alpha
\simeq 0.05$ in Fig.~\ref{fig:boxsz_mhd} for the shearing box
simulation, whereas $\alpha=0.035$ for model G5 because of the
differing magnetic field topologies and strengths.  The density
fluctuations for the shearing box model give $\langle \delta \varrho /
{\varrho}\rangle =0.17 $, whereas $\langle \delta \varrho /
{\varrho}\rangle =0.13$ for model G5.

\begin{table*}
  \caption{Global simulation parameters and results.\label{table1}}
  \begin{tabular}{@{}lllllllllcl@{}}
  \hline Simulation & $\rin$/$\rout$ & $\phi_{\rm max}$ &$Z_{\rm
  min}$/$Z_{\rm max}$ & $H/r$ & $\beta$ & $\langle { \alpha} \rangle$ &
  $\langle \delta \varrho/{\varrho}\rangle$ & $\langle \delta
  \Sigma/{\Sigma}\rangle$ & $C_{\sigma}(v_r)/[\Cs \times 10^{-3}]$ &
  $C_{\sigma}(\Delta a)/[10^{-4}]$ \\ \hline

G0/G1 & 1.5/3.5 & $\pi/2$ & $\pm 0.125$ & 0.05  & 50  & 0.035 & 0.143 & 0.109 & $8.87 \pm 0.50$ & $7.92 \pm 0.69$\\
G2 & 1.5/3.5 & $2 \pi$ & $\pm 0.125$ & 0.05  & 50   & 0.040 & 0.159 & 0.124 & $8.20 \pm 0.32$ & $12.20 \pm 0.56$  \\
G3 & 1.5/3.5 & $\pi/2$ & $\pm 0.125$ & 0.05  & 200  & 0.017 & 0.101 & 0.088 & $8.18 \pm 0.41$& $7.17 \pm 0.39$ \\
G4 & 1.5/3.5 & $\pi/2$ & $\pm 0.125$ & 0.05  & 12.5 & 0.105 & 0.266 & 0.150 & $11.30 \pm 0.70$ & $10.00 \pm 0.10$ \\
G5 & 1.0/4.0 & $\pi/2$ & $\pm 0.187$ & 0.075 & 50   & 0.034 & 0.126 & 0.094 & $4.69 \pm 0.21$ & $7.65 \pm 0.67$ \\
\hline
  \end{tabular}
\end{table*}

\begin{figure}
  \center\includegraphics[width=\columnwidth]{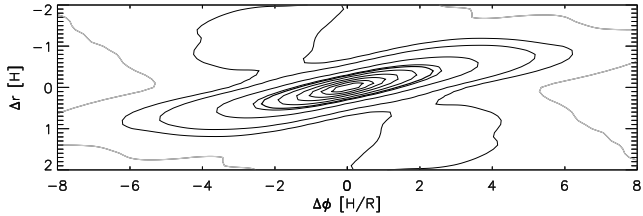}
  \caption{Contours of the two-point correlation of the surface
    density averaged over six snapshot from model G1.  The heavy
    contours represent the $e^{-1}$, and the zero level is represented
    by the grey line.}
  \label{twopointcorrelation}
\end{figure}

In Sect.~\ref{sec:MHD_turb}, it was shown that the two-point
correlation function for the density field obtained in local shearing
box simulations was highly anisotropic, with structure in the
azimuthal direction being stretched by a factor of $\sim 6$ relative
to the radial direction.  We plot the corresponding two-point
correlation function for model G1 in
Fig.~\ref{twopointcorrelation}. It is very similar to that shown in
Fig.~\ref{fig:rho_acf} for the shearing box run, indicating strong
similarities in the density structures obtained in local and global
simulations.

\subsection{Gravitational torques versus shearing box size}
\label{sec:torques}

\begin{figure}
  \begin{center}
    \includegraphics[height=0.48\columnwidth]{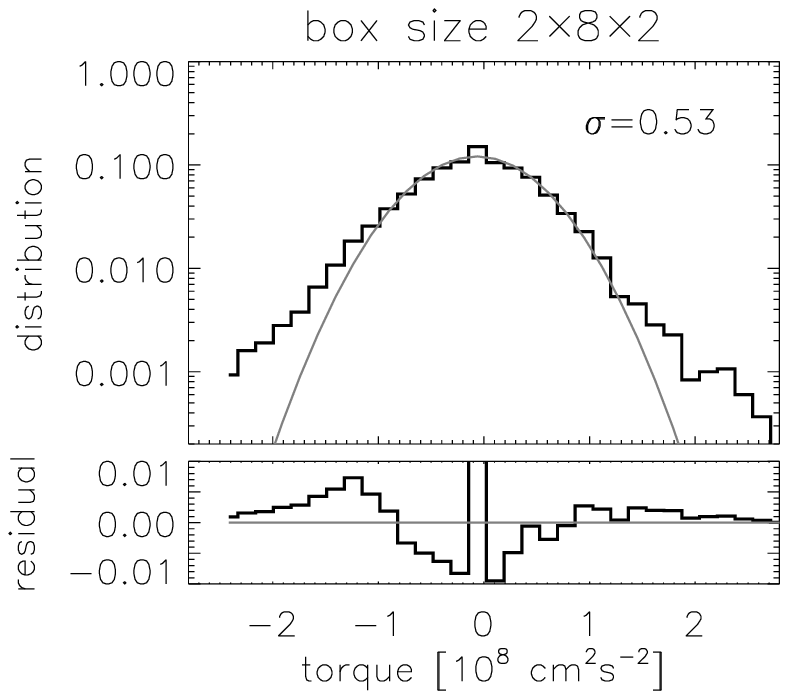}\hfill
    \includegraphics[height=0.48\columnwidth]{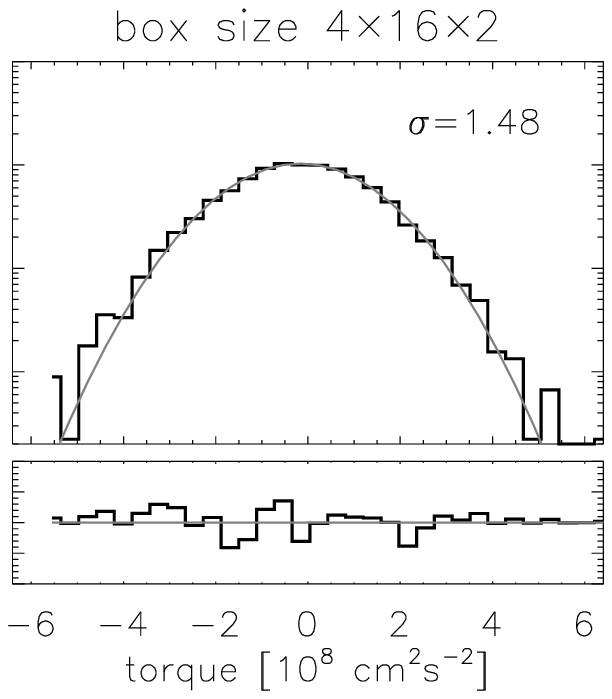}
  \end{center}
  \caption{Time averaged distribution of the gravitational torques
    acting on a set of particles. \emph{Left:} small box, the
    distribution shows excess of both large and small values (as
    typical for intermittency). \emph{Right:} large box, the histogram
    is well represented by a normal distribution centred around zero
    and with a standard deviation of $1.49\times10^8 \cm^2\s^{-2}$.}
  \label{fig:trq_dist}
\end{figure}

To illustrate the strong dependence of the disc gravity on the domain
size in local simulations, we performed sets of runs with varying
radial and azimuthal extents of the box. As a measure of the magnitude
of the stirring, we record time series of gravitational torques at
fixed positions and compute the width of the resulting distribution
function. This is exemplified in Fig.~\ref{fig:trq_dist}, where we see
that, only for large-enough boxes, the torques are consistent with a
normal distribution. Gaussian fluctuations, in turn, warrant
stochastic modelling as considered by \cite{2007Icar..192..588Y}, for
example, for the case of interactions via gas drag.
 
The intermittent distribution seen in the left panel of
Fig.~\ref{fig:trq_dist} is likely related to recurring channel modes
and the spiky nature of time series for the turbulent stresses, which
occurs when going to too narrow boxes in the radial direction. This
phenomenon, which is related to the truncation of the dominant
parasitic modes, was first observed by \cite{2008A&A...487....1B}, and
we can confirm this result with our simulations.

\begin{figure}
  \center\includegraphics[width=0.95\columnwidth]{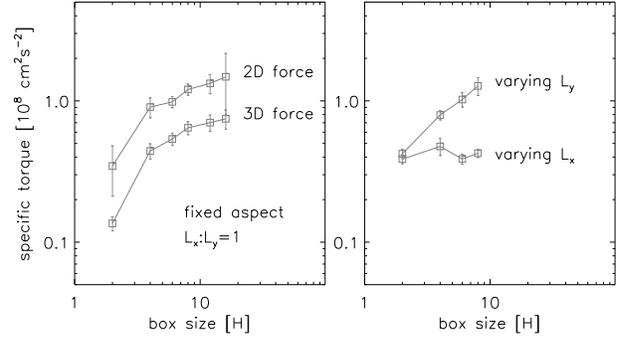}
  \caption{\emph{Left:} Specific gravitational torques vs. box size as
    determined by fitting a normal distribution
    (cf. Fig.~\ref{fig:trq_dist}).  Gravitational torques computed via
    the column density (``2D force'') are consistently enhanced by a
    factor of a roughly two, as expected from a simple geometric
    argument. \emph{Right:} Isolation of the effects due to variation
    in $L_x$, and $L_y$, respectively.}
    \label{fig:boxsz_trq}
\end{figure}

As can be seen in the left panel of Fig.~\ref{fig:boxsz_trq}, the
torques show a pronounced dependence on the box size, spanning almost
one order of magnitude. Even for $L\simgt 8H$, a weak trend towards
higher torques is visible, albeit remaining within the error bounds.

As already discussed in Sect.~\ref{sec:f_grav}, using the column
density for computing forces consistently enhances the 
torques by a constant geometric factor. For a fixed vertical extent
$L_z$, the results can be corrected accordingly.

As expected from the shape of the autocorrelation function in
Fig.~\ref{fig:rho_acf} and the arguments given in
\cite{2009MNRAS.397...52H}, the observed dependence on the box size is
primarily a dependence on $L_y$. This is illustrated in the right
panel of Fig.~\ref{fig:boxsz_trq}, where we plot the respective
dependence on $L_x$ and $L_y$ separately.

\subsubsection{Torque correlation time}
\label{sec:trq_acf}

In addition to examining the dependence of gravitational forces on the
box size, it is also important to consider the influence of the box
size on the correlation time associated with the stochastic
gravitational forces experienced by embedded bodies. A key issue here
is whether the periodicity of the shearing box, which allows
for the possibility  of waves in the flow to propagate radially 
past an embedded
planetesimal on multiple occasions prior to damping, combined
with advection due to the background shear flow, can modify the
recurrence time of the temporally varying gravitational field.

\begin{figure*}
  \begin{center}
  \includegraphics[height=0.7\columnwidth]{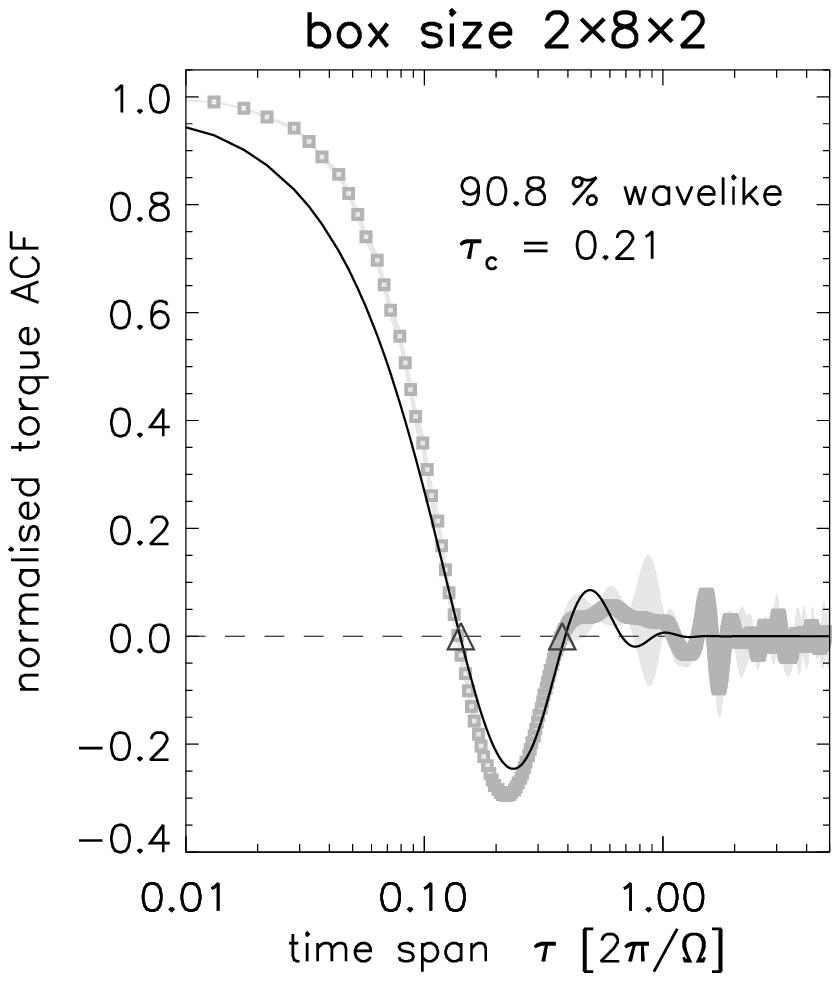}$\quad$
  \includegraphics[height=0.7\columnwidth]{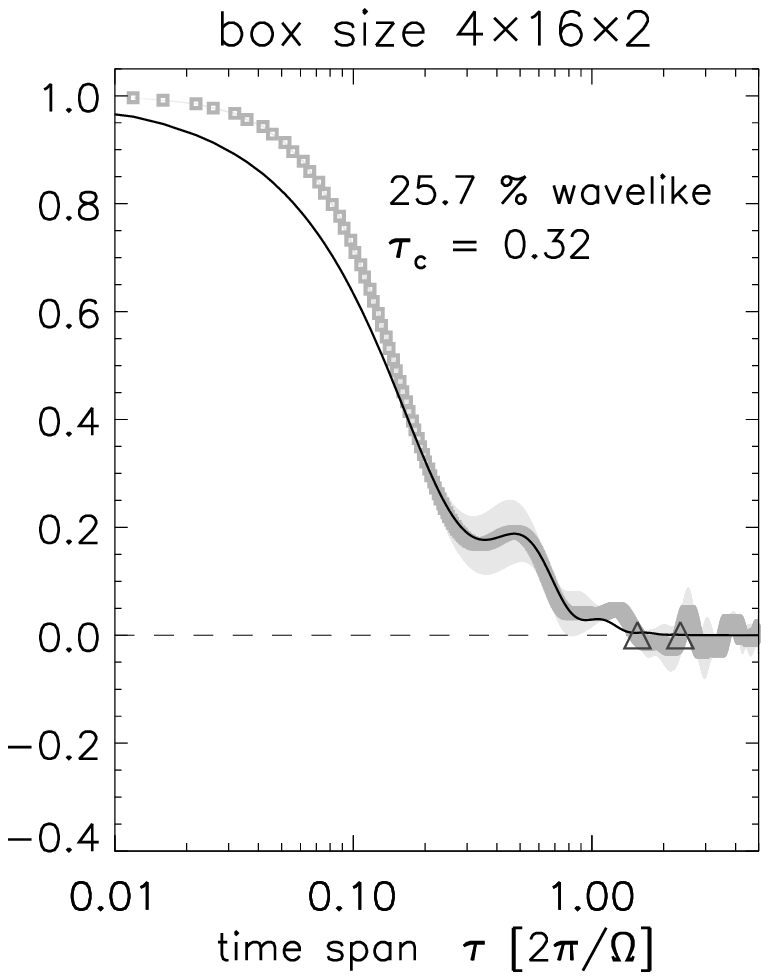}$\quad$
  \includegraphics[height=0.7\columnwidth]{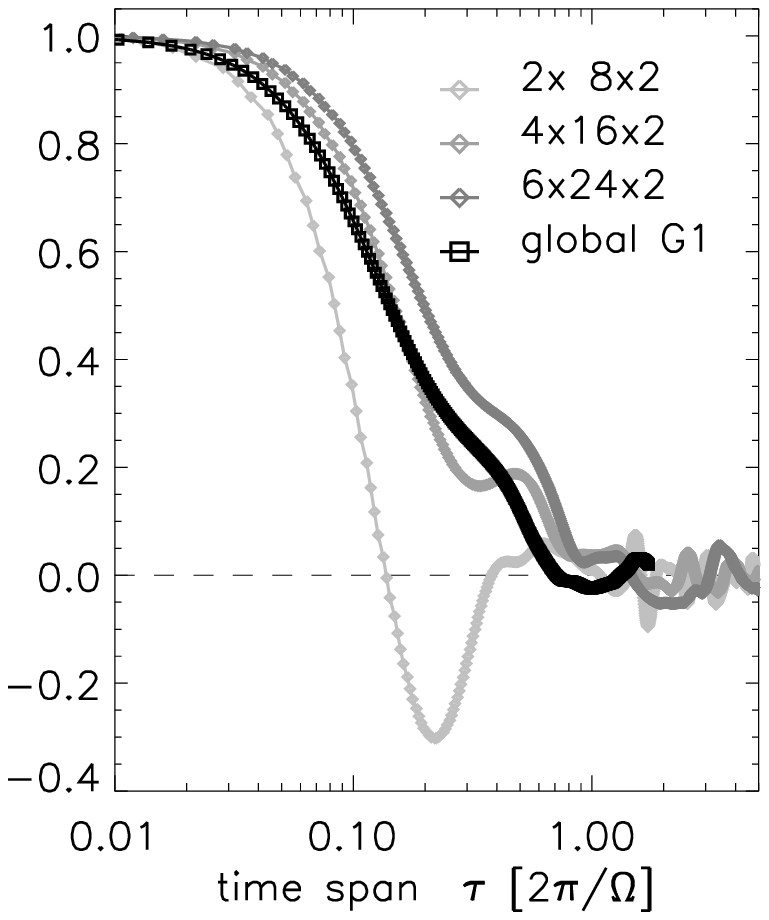}
  \end{center}
  \caption{ Box size-dependence of the torque
    autocorrelation. \emph{Left \& Centre:} Best fits according to
    Eq.~(\ref{eq:fit_acf}) (\emph{solid black line}) for a small and
    large box, respectively. The torque ACFs (\emph{dark grey lines})
    are measured at a fixed position, and error intervals
    (\emph{shaded areas}) are estimated from considering sub-intervals
    in time. The first and second zero-crossing are indicated by
    triangles. \emph{Right:} Comparison between local models and the
    global run G1 applying comparable spatial resolution. Excellent
    agreement is obtained if a large enough box-size is chosen.}
  \label{fig:ACFs}
\end{figure*}

In the previous subsection, we discussed the value of the standard
deviation of the stochastic torques obtained for runs with different
box sizes. Using the time series for the stochastic torques obtained
from these simulations, we can calculate the autocorrelation function
(ACF) for the torques as a function of box parameters. A selection of
the ACFs we obtained are plotted in Fig.~\ref{fig:ACFs}.

In the left panel of Fig.~\ref{fig:ACFs}, we show the torque ACF for a
box with size $2H\times8H$ in the horizontal direction. The plot can
be directly compared to Fig.~15 in \cite{2009ApJ...707.1233Y}, and the
results are very similar. The position of the first and second
zero-crossing as well as the relative amplitude of the negative part
of the ACF are in good agreement. \citeauthor{2009ApJ...707.1233Y}
have speculated that the observed undershoot may be responsible for
reducing the diffusion coefficient. While this might in fact be the
case, we suspect that the observed feature is probably an artifact due
to the periodic boundary conditions, limited box size and
aspect ratio.

The sinusoidal modulation of the signal in the ACF is consistent with
periodicity being introduced into the temporal evolution of the
torques caused by density waves traversing the box in the radial
direction multiple times before they are dissipated. Looking at
animations of the column density, one can identify local density
enhancements whenever two waves cross each other. These features,
which create strong torques locally, are more pronounced in smaller
boxes, and also at higher resolution (where the dissipation time is
somewhat enhanced). Studying a set of simulations, we found the
dependence on the box size and aspect ratio to be dominant, 
while the trend with resolution is rather weak
($\sim 30\%$ when increasing the resolution
by a factor of 3) -- we hence focus on this issue. To pursue a more
quantitative analysis, we fit the computed ACFs with the following
model
\begin{equation}
  S_{\Gamma}(\tau) = \left[ (1-a) + a\,\cos(2\pi\,\omega\,\tau) \right]\,
  {\rm e}^{-\tau/\tau_{\rm c}} \label{eq:fit_acf}\,,
\end{equation}
with three free parameters, namely $a$, indicating the relative
strength of the proposed sinusoidal feature, $\omega$ giving its period,
and finally the correlation time $\tau_{\rm c}$, assumed to be common
between the two components.

For the sake of simplicity, we assume that there exists only one
wave-like mode and both components of the mix decay with the same
characteristic time. While more elaborate fits (e.g. with separate
decay times) produce slightly better agreement, we found no systematic
trend in this dependence. We therefore refrain from the associated
further complication.

As one can see in the left and centre panels of Fig.~\ref{fig:ACFs},
the model produces reasonable fits for the given data. For the small
and large boxes, we infer relative amplitudes of the cosine-like
feature of $\sim90\%$ and $\sim25\%$, respectively. This illustrates
the trend with box size at a fixed aspect ratio. Contrary to our own
expectation, the periodicity is not simply related to the \emph{radial}
extent of the box. This has been found studying a set of simulations, 
keeping $L_y=8H$ fixed and progressively increasing $L_x$. Even at 
$L_x=16H$, the ACF remains unchanged and is very similar to the one 
in the left panel of Fig.~\ref{fig:ACFs}. It appears instead that
changing both the box size {\em and} aspect ratio leads to changes in
the torque correlation time, as illustrated in the centre
panel of Fig.~\ref{fig:ACFs} for a run with box dimensions
$4 H \times 16H \times 2H$. The implication appears to be that
the periodicity introduced into the run of torques versus
time, and hence into the ACF, results from a combination of wave 
propagation in radius and advection in the azimuthal direction due
to the background shear.

Having identified a periodic feature, we can now make an unbiased
estimation of the temporal correlation of the fluctuating
torques. Although unexpected from a na\"ive by-eye inspection, the
correlation times in the two cases only differ by about 50 per
cent. This shows that estimating the correlation time according
to location of the first zero crossing when the autocorrelation
function has a significant sinusoidal component can be misleading.
We suggest that as an alternative the correlation time be estimated
using a fitting formula such as given in Eq.~(\ref{eq:fit_acf}).
We see from the central panel of Fig.~\ref{fig:ACFs} that
$\tau_c = 0.32$ for our large shearing box model.

Finally, in the right panel of Fig.~\ref{fig:ACFs}, we compare the
torque ACFs of our global simulation G1 with local simulations at
comparable resolution. We see that excellent agreement is obtained
when using large-enough boxes with an elongated aspect ratio,  
indicating that the correlation time for the global simulation G1 
$\tau_c \simeq 0.32$. Interestingly, the correlation time
measured from the zero crossing point for model G1 is
$\tau_c =0.68$, about a factor of two larger than we infer
from the fitting procedure described above.

In order to check the sensitivity of model G1 to numerical
resolution, we have re-run
it at both double and quadruple resolution in the radial
and azimuthal directions, giving ($N_r \times N_{\phi}$) equivalent
to ($320 \times 2560$) and ($640 \times 5120$), respectively for
a disc which covers the full $2 \pi$ in azimuth. 
These runs have 20 and 40 cells per mean scale height in the
radial and azimuthal directions. The ACFs measured in each of 
these runs are very similar to that shown in the right panel of
Fig.~\ref{fig:ACFs}.


\subsection{Gravitational stirring versus distance in global models}
\label{global-grav-influence}
As described in Sect.~\ref{global-disc-models}, the disc models listed
in table~\ref{table1} were evolved until they had reached a statistical
steady state, prior to inserting the planetesimals.
Before discussing the evolution of planetesimals which
experience gas drag, we examine the length scales over which
gravitational stirring of planetesimals by the turbulent disc occurs.

We performed a series of simulations using the disc model G0, where we
defined a gravitational sphere of influence of varying size, $R_{\rm
cut}$, around the planetesimals (which do not experience the gas drag
force in this particular simulation) . We then examined how the radius
of this sphere of influence changed the evolution of the planetesimal
velocity dispersion. The sphere of influence in each simulation is
defined by a radius, $R_{\rm cut}$, measured in units of the local
value of $H$. Gas within this sphere of influence exerts a
gravitational force on the planetesimal. At the edge of this sphere,
the contribution to the force is tapered to zero over a distance equal
to $H$ using the hyperbolic tangent function. For example, a sphere of
influence equal to $R_{\rm cut}=2H$ allows a full contribution to the
disc force within $1H$, and beyond this the force contribution is
tapered to zero.

\begin{figure}
  
  \center\includegraphics[width=0.9\columnwidth]{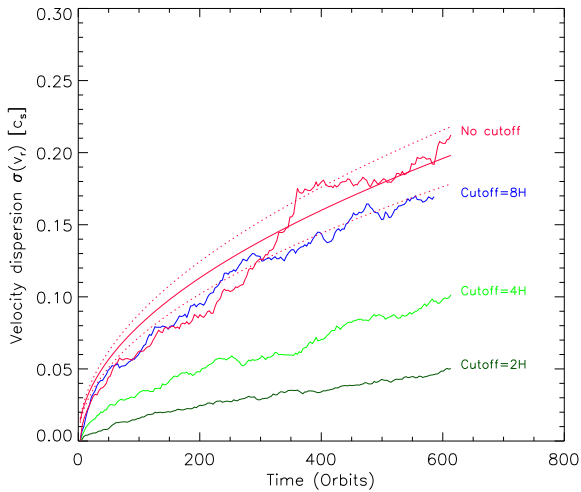}
  \caption{Evolution of the radial velocity dispersion, $\sigma(v_r)$,
    in units of the local sound speed,
    as a function of the size of the gravitational sphere of influence
    described in the text.}
  \label{vdisp-G0}
\end{figure}

Fig.~\ref{vdisp-G0} shows the evolution of the rms of the radial
velocity dispersion, $\sigma(v_r)$, as a function of $R_{\rm cut}$,
where $\sigma(v_r)$ is measured in units of the sound speed.  For this
model, $\Cs=666\ms$ at $r=2.5$ ($5\au$).  It is clear that
contributions to the gravitational force occur even beyond a cut-off
radius $R_{\rm cut}=8H$, in agreement with the calculations presented
in Sect.~\ref{sec:torques} for shearing boxes of different size. The
basic reason for this has already been explained: spiral waves
generated by the turbulence provide coherent structures which are
stretched in the azimuthal direction by the shear, and contribute
significantly to the stochastic gravitational forcing experienced by
the planetesimals.  Clearly, computational domains are required which
are large enough to capture the gravitational stirring which occurs on
scales up to 8 -- 10 scale heights, at least in the azimuthal
direction.

Examining the curve in Fig.~\ref{vdisp-G0} which corresponds to no
cut-off in the gravitational force, we see that the evolution of
$\sigma(v_r)$ can be reasonably well fitted as a random walk.  The
smooth solid line, and the dashed lines, correspond to the function
$C_{\sigma} \sqrt{t - t_0}$, with 
$C_{\sigma}=8 \pm 0.8 \times 10^{-3}$, where
the time is measured in orbits at $r=2.5$ ($5\au$).

\subsection{Evolution of the planetesimal velocity dispersion - local model}
\label{sec:vel-disp-local}
After $t_0=20$ orbits, when the turbulence driven by the
magneto-rotational instability has reached a quasi steady-state, we
disperse several swarms of test particles into the flow. For easy
reference, we label these sets as ``G'' for the particles that
experience the gas gravity only, ``D'' for particles subject to the
gas drag-force, ``G+D'' for the combined effect, and ``T'' for
massless tracer particles. While the sets G and T consist of only one
species, the sets D and G+D are composed of ten species each. Particle
radii are $R=1\m$, $2\m$, $5\m$, $10\m$, $20\m$, $50\m$, $0.1\km$,
$0.2\km$, $0.5\km$, and $1\km$, respectively. Eight particles are used
to represent each size. Due to the low number of particles, we expect
sampling errors on the order of $20-35\%$. In reality, these numbers
have to be seen as upper limits. Because our fits are based on time
histories, the effective statistical basis is probably somewhat larger
than for any given instant in time. To roughly quantify the
uncertainty due to Poisson fluctuations, we performed a lower resolved
fiducial run with 100 particles for the gravity-only set. Taking
twelve subsets of 8 particles each, we arrive at a standard deviation
of $23\%$ amongst the different realisations. For six sets of 16
particles, this number is only slightly reduced to $18\%$, justifying
the initial choice.

\begin{figure}
  \center\includegraphics[width=\columnwidth]{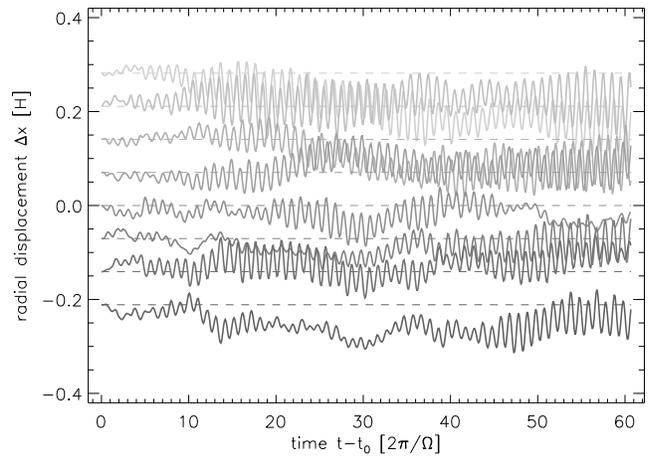}
  \caption{Exemplified temporal evolution of the radial displacement
    $\Delta x$ for a swarm of eight gravitationally excited
    particles. The initial positions have been off-set for clarity as
    indicated by the dashed lines.}
  \label{fig:xdisp}
\end{figure}

Depending on the particle size and the prevailing form of the
coupling, the time evolution of the various species is quite
diverse. Massless tracers and small particles with $\Rp\simlt 10\m$,
whose dynamics are largely controlled by gas drag, essentially follow
the turbulent flow and describe a random walk.  The larger G+D
particles, as well as the G set are coupled more weakly. For them the
orbital dynamics due to the Coriolis and tidal forces becomes
increasingly relevant. Viewed from a local perspective, the motion of
these particles can be described as epicyclic oscillations with a
modulated amplitude and a stochastically migrating guiding centre, as
seen in Fig.~\ref{fig:xdisp}.  Thus, the relevant properties of the
particles' motion can be described using two characteristic
quantities: (i) orbital eccentricity, or the amplitude of the
epicyclic motion (or equivalently the velocity dispersion relative to
a circular Keplerian orbit); (ii) semimajor axis -- the position of
the guiding centre, which evolves as a random walk as the particles
migrate from their initial locations.  In this section we examine the
eccentricity/velocity dispersion, and consider the migration in
Sects.~\ref{global-migration} and \ref{sec:schmidt}.

\begin{figure}
  \center\includegraphics[width=0.9\columnwidth]{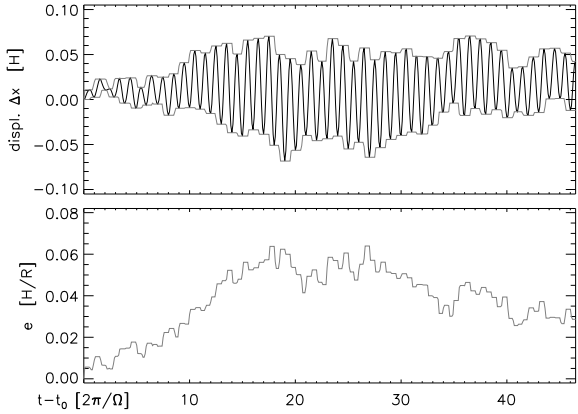}
  \caption{Illustration of the algorithm used to measure the
    eccentricity (\emph{lower panel}) of a particle moving on
    epicyclic orbits via tracking the aphelion and perihelion
    (\emph{upper panel}).}
  \label{fig:alg_ecc}
\end{figure}

To separate the stochastic motion of the guiding centre from the
epicyclic oscillation, we box car-average the velocity applying a
filter-scale equivalent of the orbital frequency. We compute the
eccentricity of the particle's orbit by tracing the position of the
aphelion and perihelion, respectively. From these, we compute
\begin{equation}
  k = \frac{R_0 + x_{\rm apo}}{R_0 + x_{\rm per}}
  \qquad \text{and} \qquad
  e = \frac{k-1}{k+1}\,,
\end{equation}
with $R_0=5\au$ the location of our local box. The tracking of the
extrema and the resulting eccentricity function are illustrated in
Fig.~\ref{fig:alg_ecc}, where we see that $e(t)$ itself follows a
random walk. This means that gravitational stirring can both excite
and damp the epicyclic motions of individual particles.  As we will
see shortly, however, the influence of the stochastic forcing on the
ensemble of planetesimals will increase the rms eccentricity amongst
the members, and therefore heat-up the ensemble as a whole.

\subsubsection{Saturation amplitudes}
\begin{figure}
  \center\includegraphics[width=0.95\columnwidth]{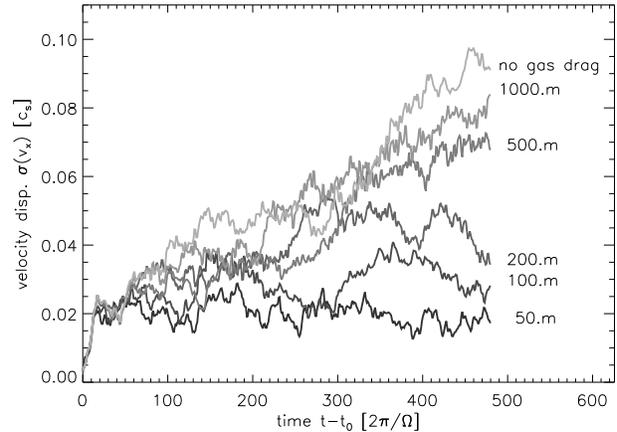}
  \caption{Evolution of the radial velocity dispersion for set 'G+D'.
    For sizes above $R\simgt1\km$, forces due to gas drag become
    gradually negligible and the evolution is described by a
    $\sqrt{t}$ behaviour.}
  \label{fig:vdist_cmp1}
\end{figure}

\begin{figure}
  \center\includegraphics[width=0.95\columnwidth]{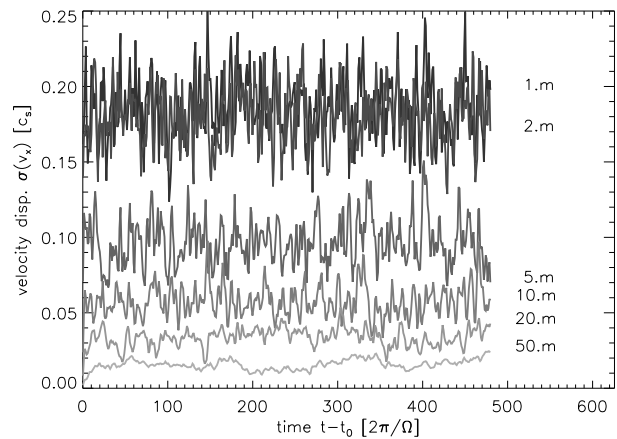}
  \caption{Same as Fig.~\ref{fig:vdist_cmp1}, but for smaller
    particles, subject only to gas drag. Note the opposite trend with
    particle radius (also cf. Fig.~\ref{fig:vsat} below).}
  \label{fig:vdist_cmp2}
\end{figure}

Needless to say that the velocity fluctuations will not grow \emph{ad
infinitum}, but will reach a saturated state once the aerodynamic
damping reaches the level of the stochastic forcing. This is
illustrated in Figs.~\ref{fig:vdist_cmp1} and \ref{fig:vdist_cmp2},
where we plot the temporal evolution of the radial velocity
dispersions for large and small bodies, respectively.

Looking at the lowermost curves in Fig.~\ref{fig:vdist_cmp1}, we see
that the random velocity of particles with $50\m \simgt \Rp \simgt
200\m$ saturates on time scales of several hundred orbits. In this
regime, the saturation amplitude increases with the particle size,
reflecting the weaker damping for larger bodies.  Because the
gravitational stirring is independent of the particle mass, smaller
bodies reach their saturated state earlier compared to heavier ones.
For larger bodies ($500\m$, $1\km$), we do not achieve equilibrium,
but after 500 orbits we note that $\sigma(v_r) \simeq 0.08 \Cs$ for
1km-sized bodies (where $\Cs=1\kms$ in this model).

Unlike for their larger counterparts, smaller bodies are predominantly
affected by gas drag and we observe the opposite trend with respect to
the saturation amplitudes (see Fig.~\ref{fig:vdist_cmp2}). At first,
it seems surprising that $\sigma(v_x)$ can be excited beyond the
turbulent velocity $v_{\rm rms}=0.13\,\Cs$ of the gas. This is,
however, only the case for the radial velocity and can be understood
from the orbital dynamics which leads to $\rms{v_x}\,\approx
2\rms{v_y}$ as characteristic of epicycles \citep[also cf. Fig.~5
in][]{2007Icar..192..588Y}.

\begin{figure}
  \center\includegraphics[width=\columnwidth]{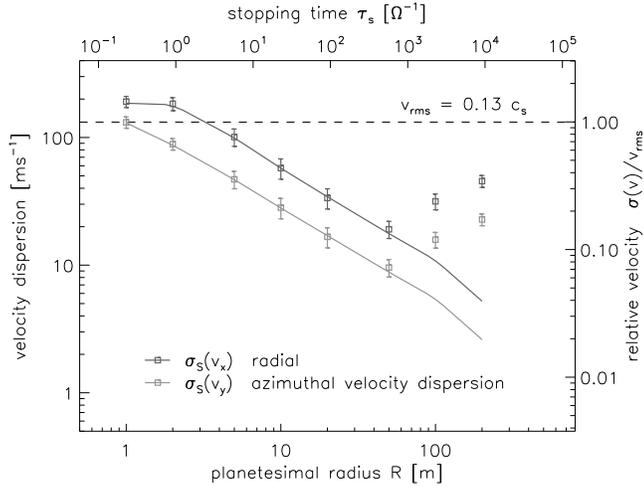}
  \caption{Saturated velocity dispersions as function of planetesimal
  radius $\Rp$, and stopping time $\tau_{\rm s}$,
  respectively. Particles which feel both the gas gravity and drag
  force (\emph{symbols}) deviate from the ones that only feel the gas
  drag (\emph{solid lines}) for sizes larger than $R\approx50\m$.}
  \label{fig:vsat}
\end{figure}

The saturation amplitudes in the velocity dispersion which arise as a
function of particle size are compiled in Fig.~\ref{fig:vsat}. With
the exception of the $\Rp=1\m$ species, which is tightly coupled to
the gas, $\sigma_{\rm\!s}(v_x)\approx 2\, \sigma_{\rm\!s}(v_y)$ as
expected from orbital dynamics. The results for set D can be readily
compared to the upper panel\footnote{As we will discuss shortly in
section~\ref{sec:schmidt}, we find a characteristic eddy time
$\tau_{\rm e}\approx \Omega^{-1}$ in our simulations such that the
standard case applies.} of Fig.~5 in \cite{2007Icar..192..588Y}, which
shows saturation amplitudes obtained from a simple turbulence
model. Considering the vast differences in the two approaches, the
results agree surprisingly well.

Considering the G+D set, the particle rms velocity takes its maximum
for $\tau_{\rm s}\simlt 1$ and falls-off to about $10$\% its peak
value at $\tau_{\rm s}\simeq 10^3\,\Omega^{-1}$. This is a somewhat
shallower decline than in the model of
\citeauthor{2007Icar..192..588Y}, where this value is reached at
$10^2$ already.  The minimum radial velocity dispersion obtained is
$\simeq 0.1 v_{\rm rms}$ (where $v_{\rm rms}$ is the turbulent
velocity dispersion of the gas). This minimum value marks the point
where the G+D set deviates from the D set (see rhs in
Fig.~\ref{fig:vsat}). This implies that objects of size $\approx 50\m$
enjoy the relative comfort of being the least affected by their
turbulent surroundings, with a velocity dispersion that corresponds to
$15-20\ms$.  Interestingly this is close to the speeds required for
catastrophic break-up of planetesimals in the 10 -- $100\m$ range
\citep{1999Icar..142....5B, 2009ApJ...691L.133S}

\begin{figure}
  \center\includegraphics[width=0.9\columnwidth]{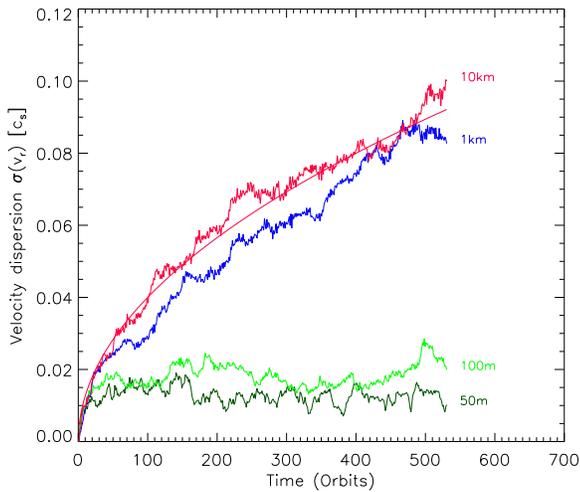}
  \caption{Evolution of the radial velocity dispersion, $\sigma(v_r)$,
    in units of the sound speed,
    for planetesimals of different size from run G5.}
  \label{vdisp-G5}
\end{figure}

For larger sizes, the velocity dispersion is seen to rise as the
particle size rises in Fig.~\ref{fig:vsat}, but the largest bodies 
we consider have not
had time to reach the equilibrium values of their velocity dispersion.
The explanation for the existence of the minimum velocity dispersion
observed in Fig.~\ref{fig:vsat} is straightforward. Smaller bodies are
tightly coupled to the gas, and so attain a velocity dispersion close
to that of the gas itself. Larger particles experience a significantly
smaller gas drag, which then contributes weakly to counterbalancing
the stirring effect of the stochastic gravitational force provided by
the disc, leading to a large velocity dispersion. Intermediate sized
particles in the 50 - $100\m$ range are sufficiently decoupled from
the gas that they experience an orbit-averaged drag force that causes
significant damping of the eccentricity growth driven by the disc
gravity.

\subsection{Evolution of the planetesimal velocity dispersion - global model}

We now examine the evolution of the velocity dispersion of
planetesimals of different size which experience the gas drag force
within global models, focusing on the radial component of this
quantity.  The key issues that we explore are the magnitude of the
velocity dispersion attained as a function of planetesimal size and
disc model parameters, and the implications for the outcomes of
collisions between planetesimals for the growth of planets.

\subsubsection{Model G5 - comparing local and global models}
The global disc model with physical parameters most similar to the
shearing box simulation is model G5. As described in
Sect.~\ref{sec:gas-drag}, the shearing box model adopted $H=0.075$,
and used Eq.~(\ref{eqn:nu-local}) and $\mu=2.4$ in defining the
strength of the gas drag force. Most of the global simulations adopted
model parameters which are slightly different from these, a fact which
was discovered after most of the simulations presented here had been
completed.  We present here, however, a global model with the same
parameters used in the shearing box run for the purpose of providing a
direct comparison. The only difference in the underlying disc models
is the choice of magnetic field topology and strength, with the
resulting $\alpha$ values being $\alpha \simeq 0.05$ for the shearing
box and $\alpha \simeq 0.035$ for the global model, which are similar
enough for a meaningful comparison to be made.  The planetesimal sizes
considered in this run were $10\m$, $20\m$, $50\m$, $100\m$, $1\km$
and $10\km$, with each size being represented by 25 particles.

The evolution of the radial velocity dispersion, expressed in units of
the local sound speed, is shown in Fig.~\ref{vdisp-G5} for
planetesimals with sizes in the range $50\m$ -- $10\km$.  Comparison
with Fig.~\ref{fig:vdist_cmp1}, which shows the same data for the
shearing box simulation, indicates that there is good agreement
between the two simulations. The $50\m$ and $100\m$ sized objects
quickly attain equilibrium values for $\sigma(v_r)$ in the range
0.01--0.02$\Cs$ (where $\Cs=1$ km s$^{-1}$ for this model), 
similar to the values obtained in the shearing box run.
After 500 orbits $\sigma(v_r) \simeq 0.08 \Cs$ for the $1\km$
sized bodies, and the $10\km$ bodies have $\sigma(v_r) \simeq 0.1
\Cs$, again in good agreement with the shearing box results.

In principle we would expect the shearing box simulations to generate
slightly larger velocity dispersions, due to the more vigorous
turbulence exhibited by that model, and the correspondingly larger
value of $\langle \delta \varrho / {\varrho}\rangle$. The fact that
the gravitational stirring is actually found to be very similar is
probably because the global simulations allow for a faster growth of
the velocity dispersion due to larger length scales being included in
the disc gravity force calculation.  More distant density
perturbations are thus able to contribute to the stochastic
gravitational field experienced by the planetesimals, providing a
small boost to the gravitational stirring.

\begin{figure}
  \center\includegraphics[width=0.9\columnwidth]{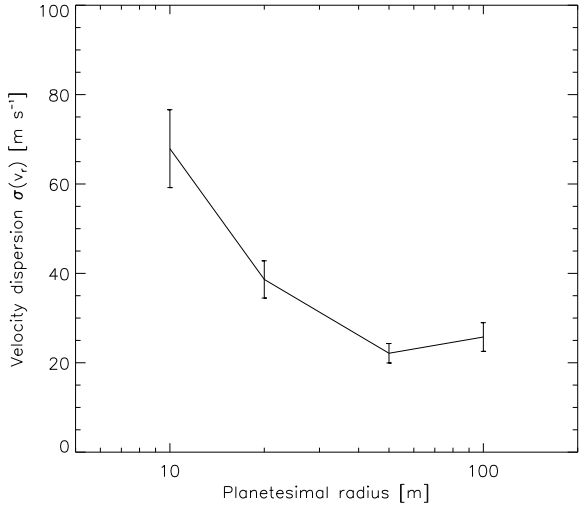}
  \caption{Saturated values of $\sigma(v_r)$, measured in metres per
    second, as a function of particle size from run G5.}
  \label{vdisp-G5b}
\end{figure}

Whereas the larger planetesimals have not achieved an equilibrium
value for $\sigma(v_r)$ by the end of the simulation, planetesimals
with sizes in the range $10\m$ - $100\m$ have. We plot the saturated
value of $\sigma(v_r)$ as a function of particle size in
Fig.~\ref{vdisp-G5b}, and in agreement with the results obtained for
the shearing box simulation, we observe that there is a minimum value
for $\sigma(v_r)$ for bodies of size $\simeq 50\m$, corresponding to
$\sigma(v_r) \simeq 20\ms$.

\subsubsection{Evolution as a function of $ \alpha$: models G1 -- G4}
\label{vdisp-versus-alpha}

\begin{figure*}
  \includegraphics[width=0.85\columnwidth]{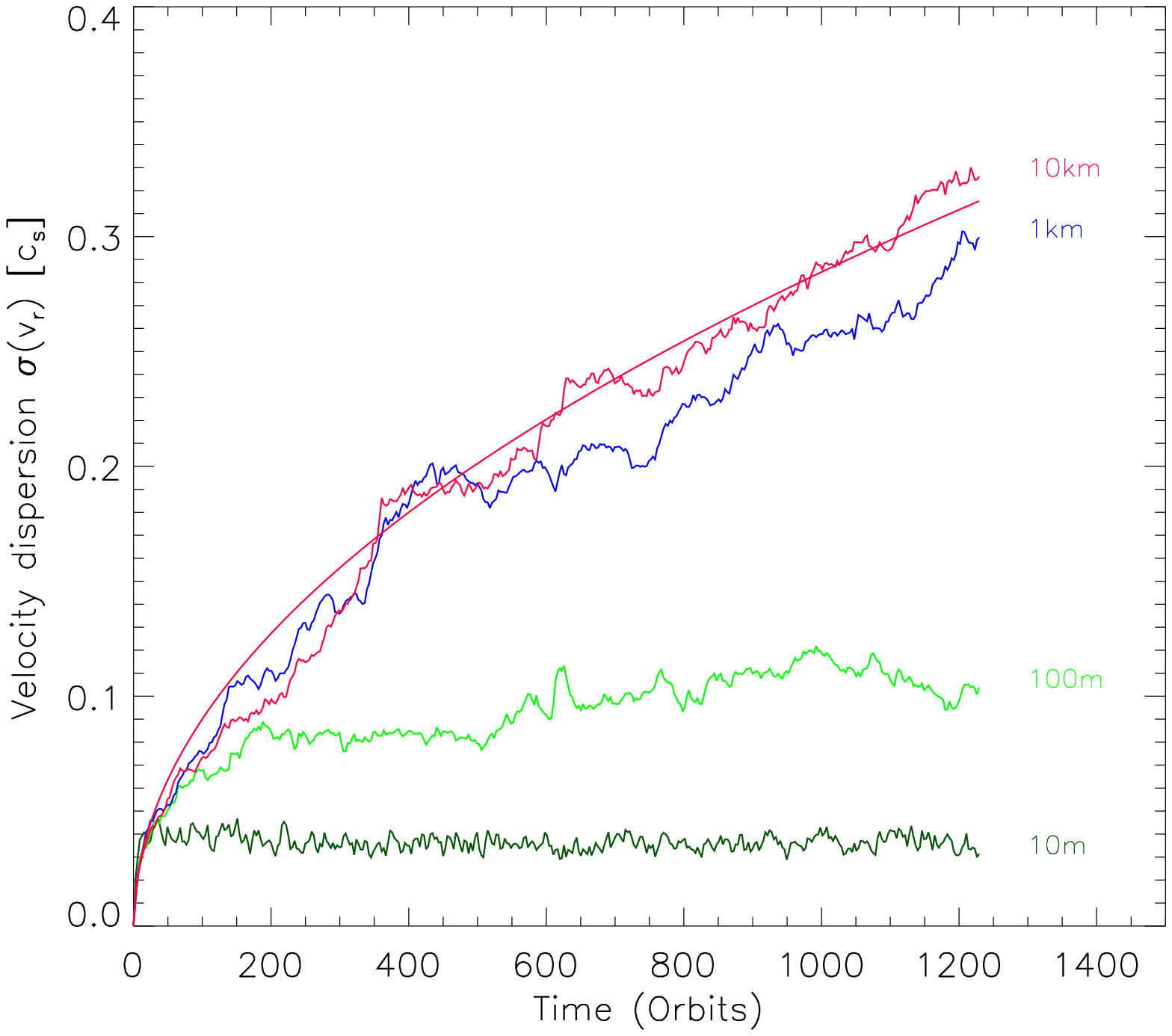}$\quad$
  \includegraphics[width=0.85\columnwidth]{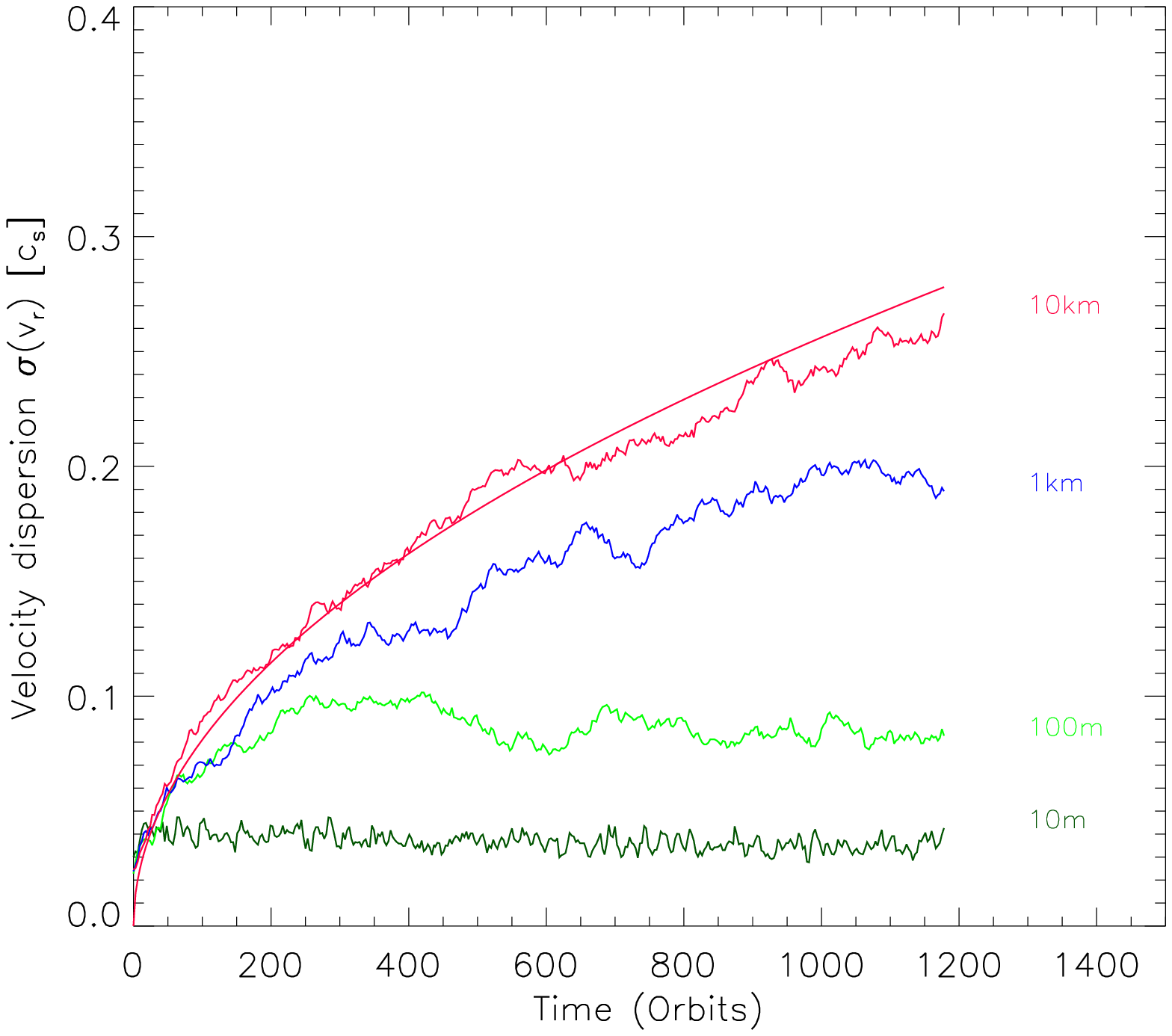}\\[1ex]
  \includegraphics[width=0.85\columnwidth]{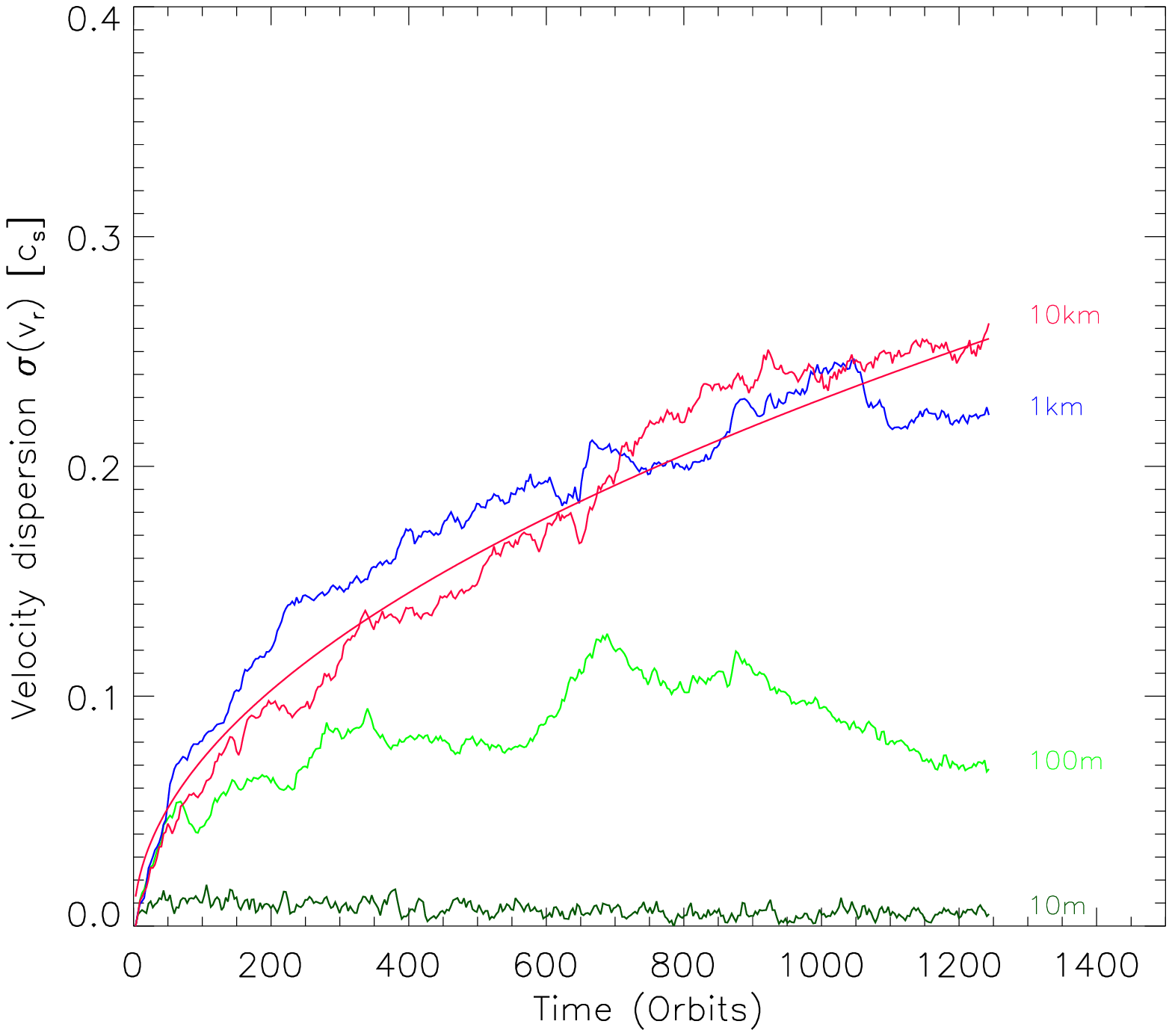}$\quad$
  \includegraphics[width=0.85\columnwidth]{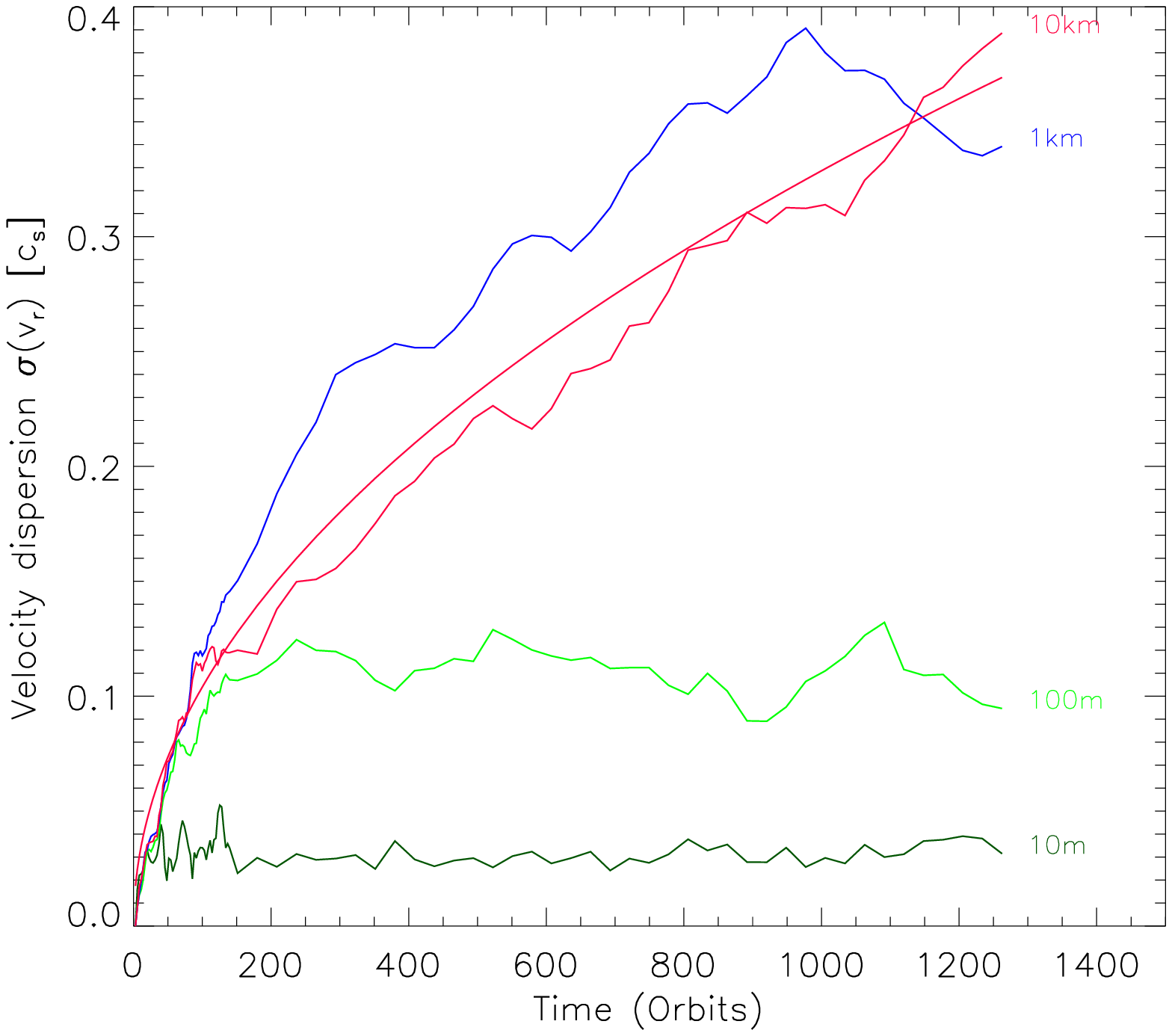}
  \caption{Evolution of the radial velocity dispersion, $\sigma(v_r)$,
    in units of the local sound speed,
    for planetesimals of different size from runs G1 (top left panel),
    G2 (top right panel), G3 (lower left panel), and G4 (lower right
    panel). Each panel also displays a fit to the random walk
    behaviour, as described in the text.}
    \label{vdisp-G1-G4}
\end{figure*}

We now consider the evolution of the velocity dispersion as a function
of the turbulent strength, as measured by ${\alpha}$, by presenting
the results from models G1, G2, G3 and G4.  As shown in
table~\ref{table1}, the value of $\alpha$ was modified by changing the
strength of the net toroidal magnetic field in the initial
conditions. Models G1 and G2 differed only in the size of their
azimuthal domains ($\pi/2$ for G1 and $2\pi$ for G2), but the increase
in domain size for G2 causes a small increase in $\alpha$
(presumably due to the presence of low-order MRI modes).  We note
that the disc parameters and values used in computing the gas drag
strength were different in these models compared with run G5, and so
these runs are not directly comparable with that one.

The evolution of $\sigma(v_r)$ for models G1 - G4 are presented in
Fig.~\ref{vdisp-G1-G4} for planetesimals of size $10\m$, $100\m$,
$1\km$ and $10\km$. Overall, the evolution of $\sigma (v_r)$ is found
to be a weak function of $\alpha$ (we find an approximate scaling
$\sigma(v_r) \propto \alpha^{0.20}$, see below), and the evolution of
$\sigma ({v_r})$ for models G1 and G2 are in good agreement.  Although
$\alpha$ (and $\langle \delta \Sigma / {\Sigma}\rangle$) is larger in
model G2, we find that the velocity dispersion increases at a slightly
slower rate for the larger planetesimals than found in model G1.  The
difference, however, is well within the $\sqrt{N}$ variations expected
for the low numbers of particles used.

In comparison with run G1, G3 shows slower growth of $\sigma(v_r)$
for the larger planetesimals, and a smaller saturated value of
$\sigma(v_r)$ for the $10\m$ sized bodies, as expected given the
smaller value of $\alpha$. Run G4 shows correspondingly faster growth,
and larger saturated values, of $\sigma(v_r)$ due to the larger value
of $\alpha$. Each of the plots in Fig.~\ref{vdisp-G1-G4} show fits to
the data for the $10\km$ bodies, assuming a functional form
$\sigma(v_r) = C_{\sigma}(v_r) \sqrt{t}$. The values of
$C_{\sigma}(v_r)$ for each model are tabulated in
table~\ref{table1}. Fitting the data for $\alpha$ and
$C_{\sigma}(v_r)$ listed in table~\ref{table1}, using an expression of
the form $C_{\sigma}(v_r)=K_{v_r} \alpha^q$, leads to a best-fit
solution with $q=0.20$ and $K_{v_r}=1.64 \times 10^{-2}$.  We use this
fit to the data in Sect.~\ref{saturation-v-alpha} below, where we
discuss the expected saturation value of $\sigma(v_r)$ for $1\km$ and
$10\km$ sized planetesimals as a function of $\alpha$.

\subsubsection{Saturation values of $\sigma(v_r)$ for 
  1km and 10km planetesimals}
\label{saturation-1km-10km}

Our simulations have not run for sufficient time for $\sigma(v_r)$ to
reach its equilibrium value for the $1\km$ and $10\km$ sized bodies.
Assuming that saturation is reached when the stochastic gravitational
forcing is balanced by gas drag damping, we can estimate the
saturation values by equating the forcing and damping time scales.
Working in terms of the orbital eccentricity (where $e \simeq v_{\rm
disp}/v_{k}$, with $v_{\rm disp} \equiv \sigma(v_r)$ and $\vK$ being
the Keplerian velocity), and using the expression $v_{\rm
disp}=C_{v_r} \sqrt{t}$ we can write
\begin{equation}
  \tau_{\rm grow} = \frac{e}{de/dt} = \frac{2 e^2 \vK^2}{C_{\sigma}(v_r)^2},
  \label{tau-grow}
\end{equation}
where $\tau_{\rm grow}$ is the eccentricity growth time.  The damping
time for the velocity dispersion can be estimated simply from the
ratio of the momentum associated with the velocity dispersion and the
gas drag force \citep{2008ApJ...686.1292I}
\begin{equation}
  \tau_{\rm damp} = \frac{2 m_{\rm p} v_{\rm disp}}{C_D \pi \Rp^2
  \varrho v_{\rm disp}^2}.
  \label{tau-damp}
\end{equation}
Equating expressions (\ref{tau-grow}) and (\ref{tau-damp}),
and writing the planetesimal mass in terms of its radius
and internal density, $\varrho_{\rm p}$, leads to the following
expression for the equilibrium velocity dispersion
\begin{equation}
  v_{\rm disp} = \left(\frac{4 \varrho_{\rm p} \Rp
    C_{\sigma}(v_r)^2}{3 C_D \varrho} \right)^{1/3}.
  \label{vdisp-equil}
\end{equation}
Noting that $\Cs=666\ms$ at $5\au$ in our global disc models G1--G4,
and expressing $C_{\sigma}(v_r)$ in S.I. units, we obtain
$C_{v_r}=3.15 \times 10^{-4}$ for model G1. This leads to estimates of
the equilibrium velocity dispersion of $v_{\rm disp}=765\ms$ for
$10\km$ bodies (approximately the sound speed), and $v_{\rm
disp}=356\ms$ for $1\km$ bodies with $\varrho_{\rm p}=3\g\perccm$ at
$5\au$ in a disc with $\Sigma=150\g\cm^{-2}$ and $H/r=0.05$.  These
values are clearly very much in excess of the velocities required for
catastrophic disruption of $1\km$ and $10\km$ sized planetesimals,
being many times larger than the escape velocities, $v_{\rm esc}$,
from these bodies ($v_{\rm esc} \simeq 12$ m s$^{-1}$ for a $10\km$
body with $\varrho_{\rm p} = 3\g\perccm$).  Extrapolating forward in
time, the time required to reach the equilibrium value for $v_{\rm
disp}$ for the $\Rp=10\km$ planetesimals is $\simeq 1.75 \times
10^5\yr$, comparable to the runaway growth time scale at $5\au$.

\subsubsection{Saturation of $\sigma(v_r)$ as a function of $\alpha$}
\label{saturation-v-alpha}
\begin{figure}
  \center\includegraphics[width=0.9\columnwidth]{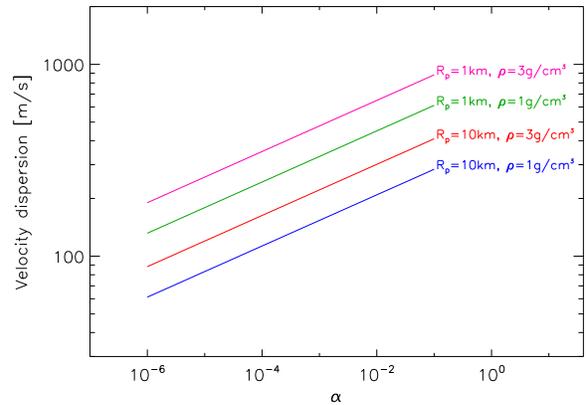}
  \caption{Variation of the saturated velocity dispersion as a function
    of the turbulent $\alpha$ and the planetesimal density.}
  \label{vdisp-v-alpha}
\end{figure}

Using the expression $C_{\sigma}(v_r) = K_{v_r} \alpha^{0.20}$
discussed in Sect.~\ref{vdisp-versus-alpha}, in conjunction with
Eq.~(\ref{vdisp-equil}), we can estimate the value of $\alpha$ which
leads to a particular value of $v_{\rm disp}$ for a particular size of
planetesimal
\begin{equation}
  v_{\rm disp} = \left(\frac{4 \varrho_{\rm p} \Rp K_{v_r}^2
    \alpha^{0.4}} {3 C_D \varrho}\right)^{1/3}.
\label{vdisp-alpha}
\end{equation}
Using the results from model G1, and working in S.I. units, we obtain
$K_{v_r}= 6.16 \times 10^{-4}$. Values of $v_{\rm disp}$ as a function
of $\alpha$ are plotted in Fig.~\ref{vdisp-v-alpha} for planetesimal
sizes $1\km$ and $10\km$, and for planetesimal internal densities
$\varrho_{\rm p}=1$ and $3\g\perccm$. It is clear that even with a small
value of $\alpha = 10^{-6}$, equilibrium velocity dispersions are in
the range $60 \le v_{\rm disp} \le 110\ms$, somewhat larger than the
escape velocities and catastrophic disruption velocities of these
bodies.  Evidently a protoplanetary disc needs to be very quiescent
near its midplane in order to allow for runaway growth to proceed, and
to prevent the catastrophic breakup of colliding planetesimals in the
1 - $10\km$ size range.

There are obviously a number of caveats contained within the simple
arguments presented above. In a disc in which the density is
vertically stratified (with or without a dead zone), the dependence of
$v_{\rm disp}$ on $\alpha$ may be steeper if the stochastic forcing of
planetesimals depends on the magnitude of local density rather than
surface density perturbations. It is also possible that mass accretion
through the disc may be generated by the winding up of net radial
fields in a disc with a dead zone \citep{2008ApJ...679L.131T},
breaking the link between $\alpha$ and $\rms{\delta\varrho/\varrho}$,
which the above arguments rely upon. Confirmation of the result
obtained above will be sought in a forthcoming paper in which we
examine planetesimal dynamics in vertically stratified discs with dead
zones.

\subsection{Radial migration of planetesimals - global models}
\label{global-migration}
We now consider the radial drift of the planetesimals, arising both
from the effects of gas drag and from the action of the stochastic
gravitational torques. The radial drift of large planetesimals is
potentially an important effect during planet formation, as it may
lead to the delivery of material to regions which have suffered from
depletion, and may also enhance the delivery of volatiles to the inner
system from beyond the snowline. Radial drift due to gas drag is not
present in the local shearing box simulations, since there is no
radial pressure gradient to generate sub-Keplerian velocities in the
gas. We therefore begin our discussion by examining the results of the
global simulations.

\begin{figure*}
  \includegraphics[width=80mm]{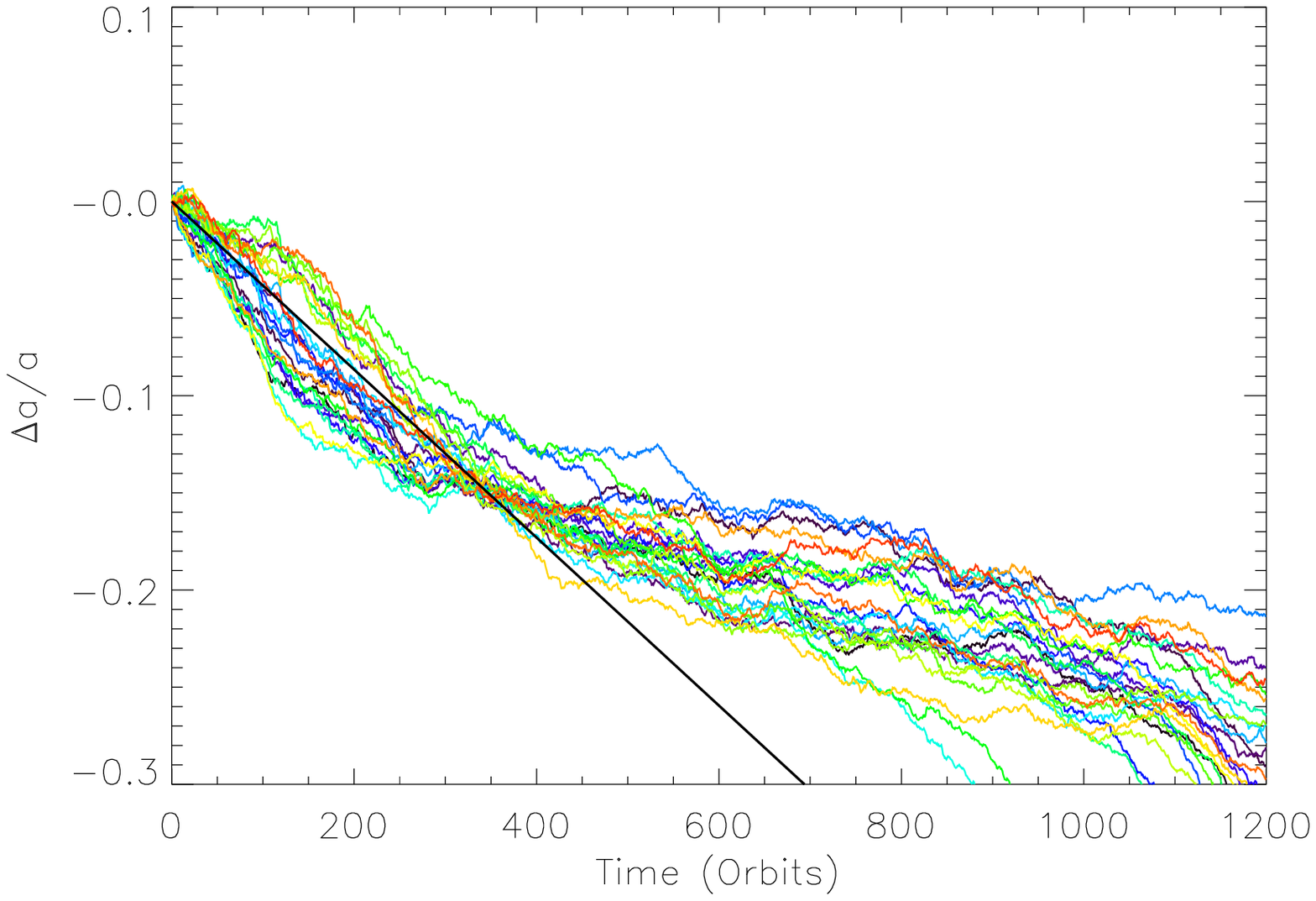}$\quad$
  \includegraphics[width=80mm]{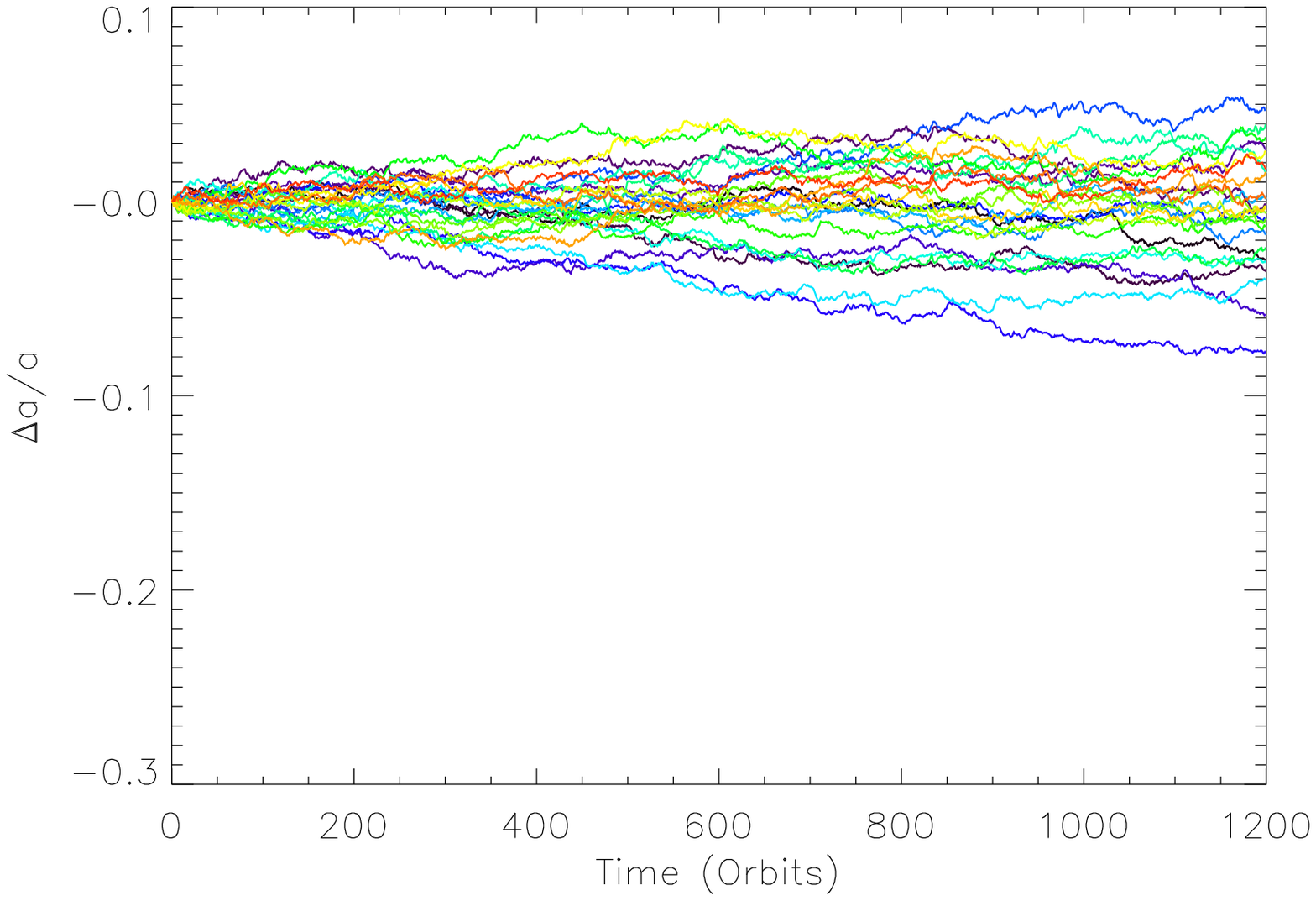}\\[1ex]
  \includegraphics[width=80mm]{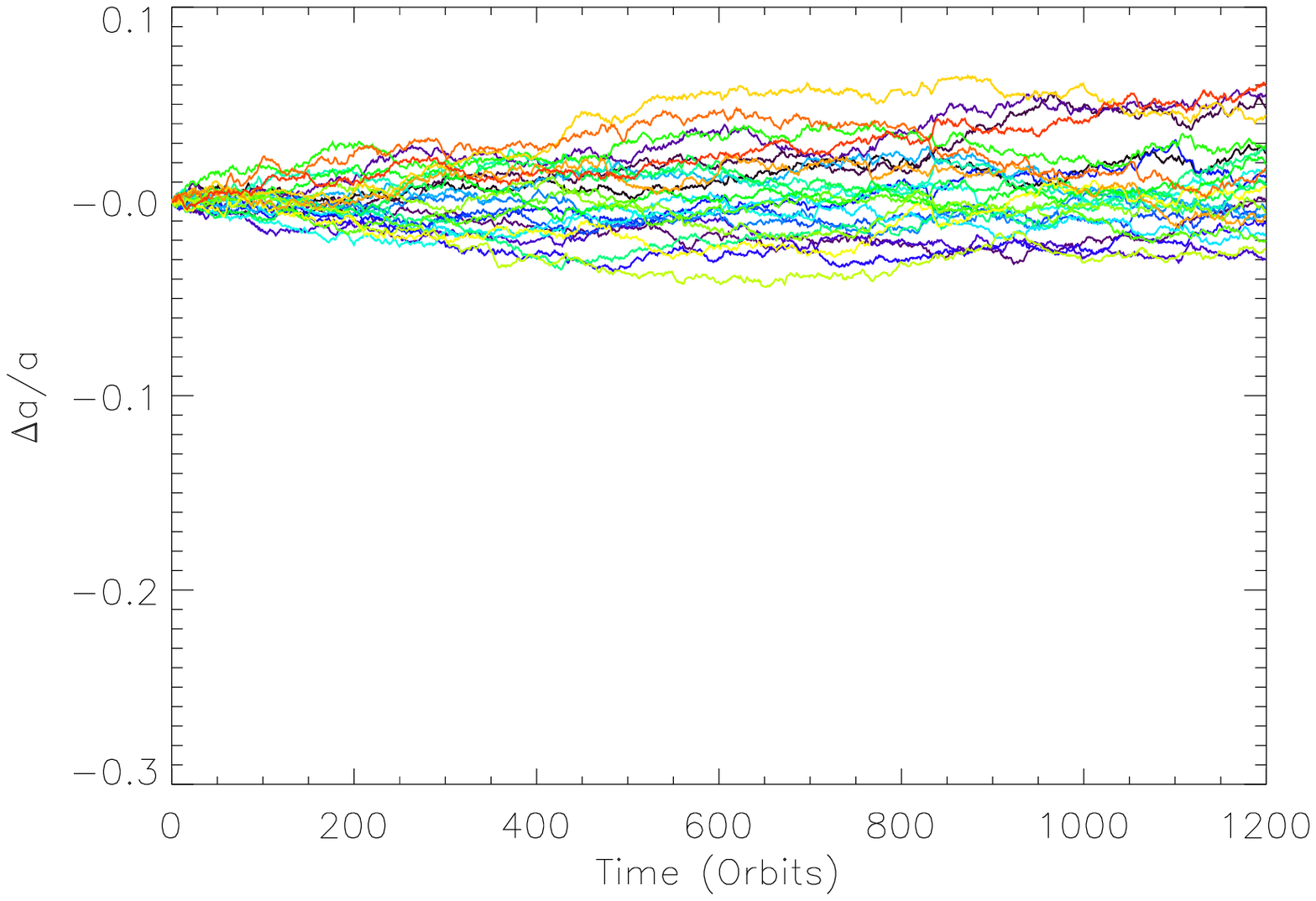}$\quad$
  \includegraphics[width=80mm]{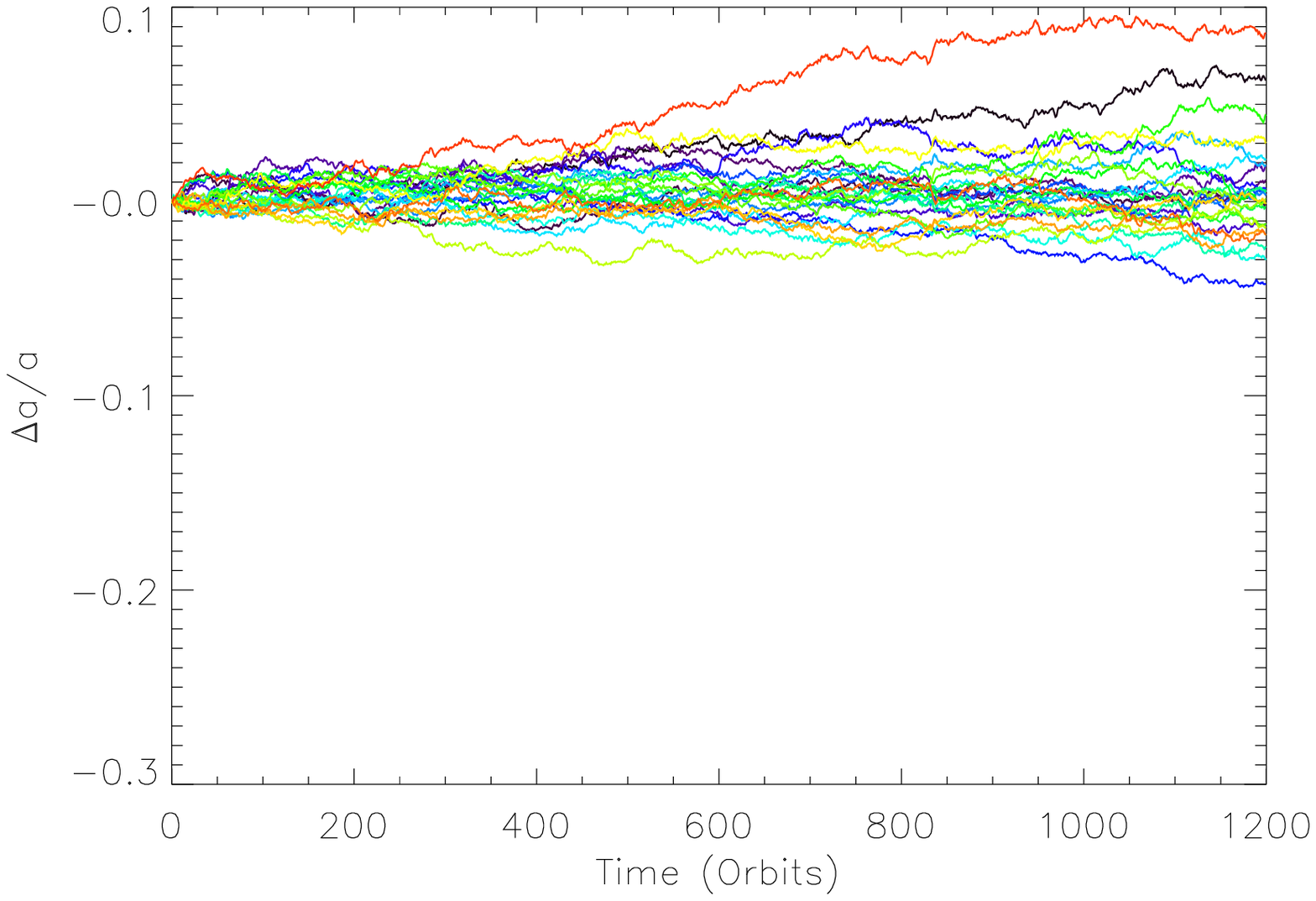}
  \caption{Evolution of the semimajor axis for all planetesimals in
    run G1. The top left panel displays data for the $10\m$
    planetesimals, along with the theoretically expected drift within
    a laminar disc (\emph{solid black line}). The top right panel
    corresponds to the $100\m$ planetesimals. The lower left panel
    refers to the $1\km$ planetesimals, and the lower right panel
    shows the $10\km$ bodies.}
  \label{migration-G1}
\end{figure*}

In a sub-Keplerian disc, the gas drag force given by
Eq.~(\ref{eqn:stokes}) will lead to a loss of angular momentum and
radial drift of planetesimals at a rate which depends on their size
\citep{1977MNRAS.180...57W}.  The time over which the gas drag removes
an amount of specific angular momentum equal to $\Delta \, j$ from an
embedded planetesimal is given by
\begin{equation}
  \tau_{\rm drag}= \frac{8 \Rp \varrho_{\rm p}}{3 C_D \varrho}
  \frac{\Delta j}{r_{\rm p} v_{pg}^2},
  \label{tau-drag}
\end{equation}
where $\varrho_{\rm p}$ is the density of the planetesimal material
(here assumed to be 3 g/cm$^3$).  Superposed on this inward drift are
the stochastic torques experienced by the planetesimals in a turbulent
disc, which will cause diffusion of planetesimal semimajor axes (as
well as contributing to the eccentricity evolution). The effective
diffusion coefficient associated with the diffusion of planetesimal
angular momenta can be approximated by
$D_j=\sigma_T^2 \tau_{\rm corr}$, where $\sigma_T$
is the standard deviation of the distribution of torques (assuming
Gaussian statistics), and $\tau_{\rm corr}$ is the correlation time
associated with the stochastic torques \citep{2006ApJ...647.1413J}.
The time scale over which diffusion will change the specific angular
momentum of a typical planetesimal by an amount equal to $\Delta j$ is
\begin{equation}
  \tau_{\rm diff} = \frac{(\Delta j)^2}{D_j}.
  \label{tau-diff}
\end{equation}

We expect that the effects of radial drift and diffusion will be
comparable when eqs.~(\ref{tau-drag}) and (\ref{tau-diff}) are equal,
and this is the time scale over which the drag-induced inward drift of
planetesimals of size $\Rp$ should just become apparent against the
isotropic diffusion generated by the turbulence.  The random walk
associated with the stochastic torques means that the planetesimal
size, $\Rp$, for which inward drift just becomes apparent depends on
the magnitude of the change of angular momentum, $\Delta j$. Larger
values of $\Delta j$ imply longer evolution times, such that the net
inward drift of larger planetesimals, which are increasingly
impervious to gas drag, eventually becomes apparent.

\begin{figure}
  \center\includegraphics[width=0.9\columnwidth]{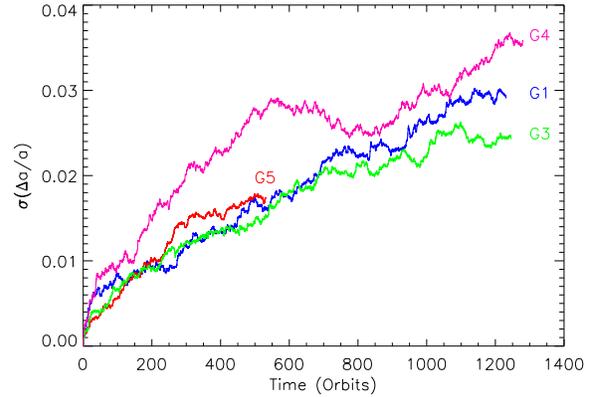}
  \caption{Evolution of the rms change in semimajor axis for $10\km$
    sized planetesimals, where each line is annotated with the run
    label.}
  \label{migration-G1-and-G4}
\end{figure}

Results from model G1 are shown in Fig.~\ref{migration-G1}, which
shows the relative change in semimajor axis versus time for all
planetesimals, and it is clear that the $\Rp=10\m$ bodies undergo gas
drag-induced inward migration, whereas the larger bodies are dominated
by stochastic migration.  The rms relative change in semimajor axis,
$\sigma(\Delta a/a)$, for the $\Rp=10\km$ planetesimals is shown in
Fig.~\ref{migration-G1-and-G4}.

\begin{figure}
  \center\includegraphics[width=0.8\columnwidth]{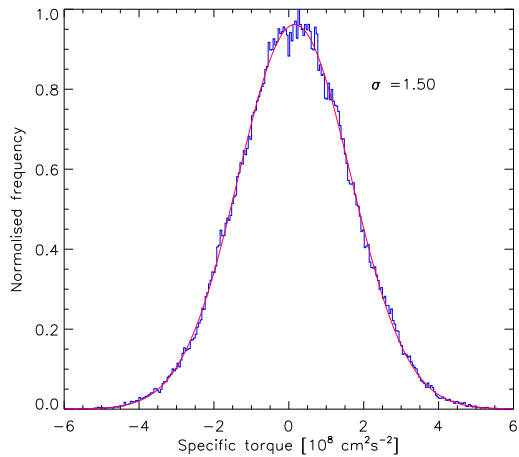}
  \caption{Distribution of torques experienced by the $10\km$ sized
        planetesimals in run G1, averaged over all planetesimals. The
        standard deviation obtained for the distribution $\sigma=1.5
        \times 10^8$ cm$^2$ s$^{-1}$.}
  \label{torques-G1}
\end{figure}

Small changes to the specific angular momentum and semimajor axis of a
body are related according to
\begin{equation}
  \frac{\Delta j}{j} = \frac{1}{2} \frac{\Delta a}{a}.
  \label{delta-j}
\end{equation}
Using Eqs.~(\ref{tau-diff}) and (\ref{delta-j}), we can estimate how
far we expect large planetesimals to have diffused in semimajor axis
during the 1200 orbits of run time in model G1. The distribution of
torques, averaged over all 10 km-sized planetesimals, is presented in
Fig.~\ref{torques-G1}, and the standard deviation of this torque
distribution is found to be $\sigma_T = 1.5 \times 10^{8}\cm^2$
s$^{-2}$ (equivalent to $3.46 \times 10^{-5}$ in code units).  The
correlation time, $\tau_{\rm corr}$, was estimated to be $\tau_{\rm
corr}=0.32$ orbits from the exponential decay of the ACF shown in
Fig.~\ref{fig:ACFs} in Sect.~\ref{sec:trq_acf}.  The diffusion
coefficient, $D_j$, expressed in code units is given by $D_j = 9.51
\times 10^{-9}$, which leads to the prediction that a typical
planetesimal located at $r=2.5$ (5 AU) in model G1 will diffuse such
that $\Delta a/a =0.021$. Comparing this with the value of
$\sigma(\Delta a/a)$ in Fig.~\ref{migration-G1-and-G4}, we see that
the simulation resulted in $\sigma(\Delta a/a) \simeq 0.03$ after 1200
orbits, in reasonable agreement with the value predicted by the
diffusion coefficient.  The fact that the estimated amount of
diffusion is too small by a factor of $\sim \sqrt{2}$ compared to the
simulation result indicates that the approximation to the diffusion
coefficient given by $D_j=\sigma_T^2 \tau_{corr}$ is too small by a
factor of $\simeq 2$.  This possibly arises because of the ambiguity
discussed in Sect.~\ref{sec:trq_acf} regarding the definition of the
correlation time, but may also be affected by sampling errors that
arise from using only 25 particles per size bin, leading to Poisson
errors at the $\sim 20$ \% level.

We can now examine if the gas drag-induced radial drift of the
$\Rp=10\m$ bodies observed in Fig.~\ref{migration-G1}, and the
stochastic migration of all larger planetesimals, is expected.
According to Eq.~(\ref{delta-j}) and Fig.~\ref{migration-G1-and-G4},
stochastic migration causes a relative change in specific angular
momentum $\Delta j/j=0.015$.  Putting this value into
Eq.(\ref{tau-drag}) tells us that planetesimals of size $\Rp \simeq
25$ metres should radially drift due to gas drag and diffuse by a
similar distance. Planetesimals which are smaller than this should
show strong inward drift, and larger bodies should have their
evolution dominated by diffusion.  The relative change in semimajor
axes for all planetesimals in run G1 plotted in
Fig.~\ref{migration-G1} show good agreement with this expectation.

Changes in the mean drag-induced radial drift rate due to the
turbulence can only be discerned for planetesimals of size $\Rp=10\m$
in Fig.~\ref{migration-G1}, where the straight black line plotted in
the top left panel shows the trajectory of a planetesimal embedded in
a laminar disc.  Up to a time of approximately 300 orbits, the mean
drift rates of the planetesimals in the turbulent disc closely match
the trajectory in the laminar disc. At later times, however, the
trajectories diverge, with the drift rates for the planetesimals in
the turbulent disc decreasing by about a factor of three. This is not
due to the effects of stochastic forces, which earlier on in the
evolution are seen to cause a dispersion of the trajectories around a
mean which matches the laminar case closely. Instead, variations in
the effective $\alpha$ stress parameter as a function of radius cause
radial structuring of the disc, such that peaks and troughs in the
mean surface density arise. The slowing down of the radial drift
observed in Fig.~\ref{migration-G1} arises when the planetesimals
enter a region of the disc in which the magnitude of the radial
pressure gradient decreases. Thus, we observe that significant changes
to the radial drift of planetesimals do not occur due to stochastic
forcing, but may arise because of radial structuring of the disc due
to spatial variations in turbulent stresses, which persist over the
run times of the simulations.

\begin{figure}
  \center\includegraphics[width=0.9\columnwidth]{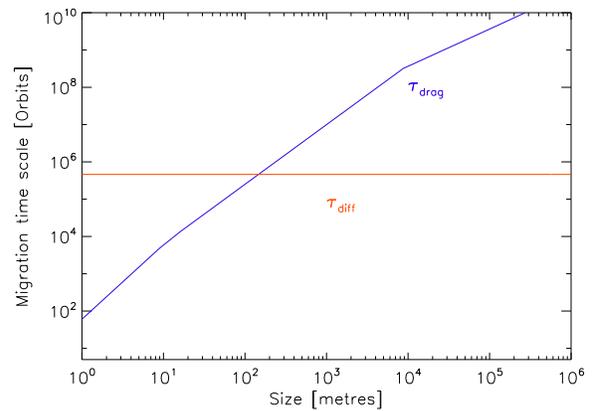}
  \caption{Variation of $\tau_{\rm drag}$ and $\tau_{\rm diff}$ as a
       function of planetesimal size. The underlying disc model is the
       same as in run G1.}
  \label{migration-v-size}
\end{figure}

If we adopt a typical disc life-time of $5\Myr$
\citep{2001ApJ...553L.153H}, then we can estimate the amount of radial
diffusion for large ($\Rp \ge 1\km$) planetesimals during the planet
forming epoch. Adopting a diffusion coefficient, $D_j$, which gives
agreement with the amount of radial diffusion observed for 10 km-sized
planetesimals in Fig.~\ref{migration-G1-and-G4}, Eq.(~\ref{tau-diff})
predicts $\Delta j/j = 0.3$, corresponding to a 50 \% change in the
semimajor axis of a typical planetesimal located initially at 5 AU.
Considering the relative contributions to radial drift of solid bodies
from both gas drag-induced migration, and diffusion due to the
stochastic gravitational field of the disc, we have plotted the
evolution times $\tau_{\rm drag}$ and $\tau_{\rm diff}$ in
Fig.~\ref{migration-v-size}, assuming a 30 \% change in the specific
angular momentum of a planetesimal, using Eqs.(~\ref{tau-drag}) and
(~\ref{tau-diff}).  We can see that for evolution times of $5\Myr$,
the drag and diffusion time scales are equal for planetesimal size
$\Rp \simeq 100\m$, but drag time scales are much larger than
diffusion time scales for larger bodies such as 1 km and 10 km-sized
planetesimals.  This shows that large scale migration of large
planetesimals over evolution times of $5\Myr$ will be dominated by
diffusion and not gas drag-induced inward drift.

Strong radial mixing of planetesimals at the level of
$\Delta a / a =0.5$ would have very significant implications
for planetary formation.  In our own Solar System,
for example, radial mixing of large icy planetesimals
from beyond the snowline would substantially increase
the volatile content of the terrestrial planets, and effectively smear
out the observed apparent variation in asteroid properties as a function
of heliocentric distance. It thus seems unlikely that
such radial mixing occurred in the Solar nebula, providing strong
circumstantial evidence that the degree of turbulence present in model
G1 was not present in the Solar nebula during the formation of the
Solar System.  We explore the degree of radial mixing as a function of
turbulent strength (measured by the effective $\alpha$ parameter)
below.

\subsubsection{Planetesimal diffusion as a function of $\alpha$}
\label{sec:diffusion}

The rms value of the relative change in the semimajor axes for $10\km$
sized planetesimals for various models is shown in
Fig.~\ref{migration-G1-and-G4}. 
The discs in models G1 and G5
gave rise to very similar values of $\alpha$
and $\langle \delta \Sigma/{\Sigma}\rangle$, and we
see that the radial diffusion rates are 
also very similar.
Model G3 had a value of $\alpha=0.017$, and a correspondingly smaller
value of $\langle \delta \Sigma / {\Sigma}\rangle$, which leads to a
slower rate of diffusion. The rate of diffusion in semimajor axis is a
fairly weak function of $\alpha$ (see discussion below), such that it
is only after a time of $t \sim 600$ orbits that the divergence in the
diffusion rates can be detected for models G1 and G3 in
Fig.~\ref{migration-G1-and-G4}. Model G4 has a substantially larger
value of $\alpha=0.101$, and leads to a noticeably larger rate of
diffusion of $\Delta a/a$ in Fig.~\ref{migration-G1-and-G4}.

We have fit the data for $\sigma(\Delta a/a)$ shown in
Fig.~\ref{migration-G1-and-G4} assuming a simple random walk model
$\sigma(\Delta a/a) = C_{\sigma}(\Delta a) \sqrt{t}$, and the
coefficients $C_{\sigma}(\Delta a)$ are given in table~\ref{table1}.
If we fit the data given in table~\ref{table1} for $\alpha$ and
$C_{\sigma}(\Delta a)$ using the function $C_{\sigma}(\Delta
a)=K_{\Delta a} \alpha^q$, then the best-fit solution gives $q=0.20$
and $K_{\Delta a}=1.70 \times 10^{-3}$.  This is the same power law
dependence found for the evolution of the velocity dispersion as a
function of $\alpha$. This fit allows us to estimate the amount of
radial diffusion that we expect as a function of $\alpha$.

\subsubsection{Radial diffusion constraints on $\alpha$ in the solar nebula}
\label{sec:constrain-alpha}
As discussed above, large planetesimals experience only small changes
in their angular momenta and semimajor axes due to gas drag, assuming
that they remain on approximately circular orbits.  For example, a
$10\km$ sized planetesimal orbiting at $5\au$ in model G1, in the
absence of turbulence, would migrate a distance of $\sim 4 \times
10^{-3}\au$ in $5\Myr$. For such large bodies, evolution of the
semimajor axis is dominated by turbulence-induced diffusion,
especially if the disc is fully turbulent as in models G1 - G5.  Model
G1 predicts that a typical $10\km$ sized planetesimal orbiting at
$5\au$ will diffuse a distance $\Delta a \simeq 2.5\au$ in $5\Myr$.
If this 50 \% change in semimajor axis is applied to bodies in the
asteroid belt, then this level of diffusion is probably inconsistent
with the variation in observed properties of asteroids as a function
of heliocentric distance \citep{1982Sci...216.1405G,
2003Icar..162...10M}.  These variations would have been smeared out
substantially if such a large amount of orbital diffusion had taken
place during Solar System formation.

The survey of how the distribution of asteroidal taxonomic types
varies as a function of heliocentric distance presented by
\cite{1982Sci...216.1405G} concluded that there is a systematic
variation. For the seven taxonomic classes identified in their sample,
it was suggested that the distribution of each peaks at a different
location in the asteroid belt, with a dispersion around this peak
location of 0.5 -- 1 au. The observed correlation between asteroidal
type (assumed to relate to composition) and heliocentric distance was
interpreted as being evidence that the asteroids were formed
essentially in their observed locations, being subject subsequently to
a relatively modest amount of radial mixing at the level of $\sim 0.5
\au$. The more recent survey of \cite{2003Icar..162...10M} presents a
more complicated picture in which S-type asteroids are distributed
more uniformly throughout the asteroid belt, but are nonetheless the
dominant class in the inner and middle belt, with the more
volatile-rich C-type bodies being preferentially located in the outer
belt beyond $\simeq 3 \au$.

Radial mixing of planetesimals during planet formation is expected to
occur through gravitational interaction with planetary
embryos. \cite{1992Icar..100..307W} suggested that a population of
$\sim$ Mars-mass embryos embedded within the primordial asteroid belt
could explain the required mass depletion and radial mixing.  More
recent simulations of this effect
\citep{2001Icar..153..338P,2007Icar..191..434O} show that radial
diffusion of asteroids peaks at a value $\Delta a \simeq 0.5 \au$,
broadly consistent with the observational constraints. In a series of
related simulations, \cite{2006Icar..184...39O} examined the formation
of terrestrial planets, and estimated the rate of water delivery to
the Earth by accretion of volatile-rich material from beyond $\sim 2.5
\au$.  It was found that simulations initiated with Jupiter- and
Saturn-analogues on circular orbits could reasonably explain the
abundance of water on the Earth ($\simeq 5 \times 10^{-4}$
M$_{\oplus}$).  Substantial radial mixing of volatile-rich
planetesimals and embryos from the outer asteroid belt, due to disc
turbulence, would very likely lead to terrestrial planets which are
endowed with water and other volatiles well in excess of what is
observed.

We suggest that orbital diffusion due to turbulence which exceeds
$\Delta a \simeq 0.5 \au$ is probably inconsistent with observations,
and we now discuss which value of $\alpha$ is consistent with $\Delta
a \simeq 0.5 \au$ in the asteroid belt.  It should be noted, however,
that using the asteroid belt as a test for dynamical mixing requires
that the observed variation of properties with heliocentric distance
is primordial. The recent suggestion by \cite{2009Natur.460..364L}
that cometary material originating from further out in the Solar
System became implanted in the asteroid belt during the late heavy
bombardment means we should take care not to over interpret the
observations, or their theoretical implications.

Using the approximate fit $C_{\Delta a} = K_{\Delta a} \alpha^{0.20}$,
discussed in Sect.~\ref{sec:diffusion}, combined with $\sigma(\Delta
a/a)=C_{\Delta a} \sqrt{t}$, we can estimate which value of $\alpha$
should lead to an amount of diffusion which is consistent with the
above discussion. Assuming that the dispersion in asteroid properties
as a function of heliocentric distance is consistent with diffusion
generating $\sigma(\Delta a / a) = 0.16 (\equiv 0.5/3)$ over a
putative solar nebula life-time of $5\Myr$
\citep{2001ApJ...553L.153H}, and noting that $K_{\Delta a} = 1.7
\times 10^{-3}$ (see Sect.~\ref{sec:diffusion}), we find an upper
limit for $\alpha \simeq 5 \times 10^{-5}$.  Interestingly, this is
very similar to the midplane Reynolds stress obtained in shearing box
simulations of discs with dead zones
\citep{2003ApJ...585..908F,2007ApJ...659..729T,
2008A&A...483..815I}. It would seem that our rough estimate for the
amount of diffusion which can occur as a function of $\alpha$ provides
circumstantial evidence that the solar nebula did indeed contain a
substantial dead zone in the vicinity of the asteroid belt and giant
planet formation region. We note, however, that caution must be used
when interpreting this result. Confirmation of the variation of
diffusion rate as a function of $\alpha$ is required using simulations
which include vertical stratification and dead zones, and account
needs to be made of the fact that the nebula mass changes
significantly over a $5 \Myr$ period.  Models for the formation of
giant planets require nebula masses approximately 3 - 5 times larger
than the MMSN \citep{1996Icar..124...62P}, which will increase the
value of $\sigma(\Delta a/a)$ by a similar factor. The disc mass will
deplete over the nebula life time, so that the net effect of these
competing factors needs to be examined in more detail.

Taken together, the constraints on $\alpha$ lead to some interesting
conclusions regarding the strength of turbulence in the solar nebula.
Shearing box simulations with dead zones generate Reynolds stresses at
the midplane with $\alpha \sim 10^{-5}$.  If this stress is
accompanied by density waves and stochastic torques with amplitudes
which follow the scaling described in Sect.~\ref{sec:diffusion}, then
it becomes very difficult to see how runaway growth could have caused
rapid growth of planetary embryos {\it via} planetesimal
accretion. Indeed, our results suggest that with $\alpha = 10^{-6}$,
catastrophic disruption of $10\km$ sized bodies remains possible
(although this conclusion must await firm confirmation from more
realistic, vertically stratified simulations).  Taken at face value,
this indicates that a mechanism is required for forming large bodies
which avoids the need to go through the runaway growth phase, such as
that suggested by \citet{2007Natur.448.1022J}.  The existence of
gradients in the chemical composition of solar system bodies as a
function of heliocentric distance, however, is broadly consistent with
a nebula model in which there exists a dead zone with Reynolds stress
$\alpha \simeq 10^{-5}$.  The existence of these chemical gradients
however, is apparently inconsistent with a fully turbulent disc model
with effective $\alpha \gg 10^{-5}$.

\begin{figure}
  \center\includegraphics[width=\columnwidth]{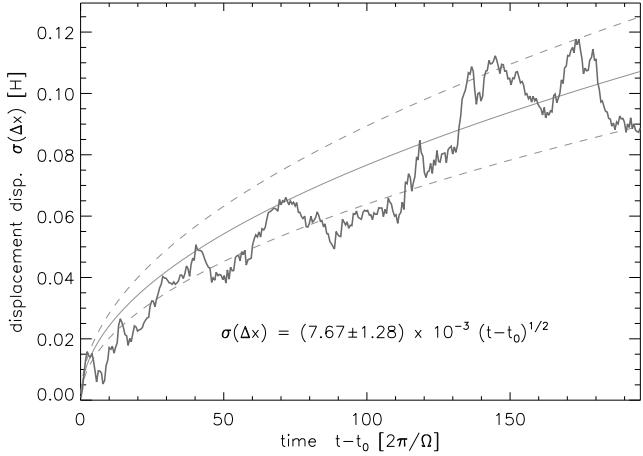}
  \caption{Dispersion in the radial displacement $\Delta x$ for a
    swarm of particles subject to the gas gravity alone. The time
    evolution follows a random walk as reasonably well described by
    the given relation.}
  \label{fig:vdist_dx}
\end{figure}

\subsubsection{Evolution of the eccentricity and $\Delta a/a$}
\label{sec:eccentricity}
For small values of the eccentricity, the relation between the rms
eccentricity, $\sigma(e)$, and the radial velocity dispersion,
$\sigma(v_r)$, may be approximated as $\sigma(e) \simeq
\sigma(v_r)/\vK$, where $\vK$ is the Keplerian velocity. In
Sect.~\ref{vdisp-versus-alpha}, we showed that for the larger
planetesimals with $\Rp=10\km$, the radial velocity dispersion may be
approximated as a random walk with $\sigma(v_r)=C_{\sigma}(v_r)
\sqrt{t}$, where $\sigma(v_r)$ is measured in units of the local sound
speed, $\Cs$, and the $C_{\sigma}(v_r)$ values are listed in
table~\ref{table1}. Similarly, we have shown in
Sect.~\ref{sec:diffusion} that the rms radial diffusion $\sigma(\Delta
a/a) = C_{\sigma}(\Delta a) \sqrt{t}$, where the $C_{\sigma}(\Delta
a)$ values are also listed in table~\ref{table1}.  Thus we expect the
ratio of eccentricity changes to semimajor axis changes,
$\sigma(e)/\sigma(\Delta a/a) = (H/r)
C_{\sigma}(v_r)/C_{\sigma}(\Delta a)$. Taking the value $H/r=0.05$ for
the models G1 - G4, we find that $ 0.34 \le \sigma(e)/\sigma(\Delta
a/a) \le 0.57$, so that changes in the relative semimajor axes are
similar to eccentricity changes for larger planetesimals that are
subject to stochastic gravitational interaction with a turbulent disc.

\subsubsection{Diffusion versus type I migration of low mass protoplanets}
\label{sec:typeI}
In addition to considering the radial drift of planetesimals due to
gas drag, and the role of stochastic torques in potentially inhibiting
this inward drift, we can also consider the effect that these
stochastic torques might have on the type I migration of low mass
planets \citep{2004MNRAS.350..849N, 2005A&A...443.1067N}.  We have
ascertained above that bodies undergoing diffusion due to stochastic
torques may change their semimajor axes by 50 \% over a disc life time
of $5\Myr$. Using the type I migration formula from
\citet{2002ApJ...565.1257T} for a disc model with properties which are
the same as model G1, we note that planets with $m_{\rm pl} /
M_{\star}=3.5 \times 10^{-7}$ will migrate a distance equal to 50 \%
of their initial semimajor axis (assumed to be $5\au$) {\it via} type
I migration over $5\Myr$. This indicates that direct competition
between type I migration and stochastic migration will only prevent
the large scale migration of planetary bodies with masses similar to
that of Mars over such nebula life times. It should be noted, however,
that the possible radial structuring of protoplanetary discs by
turbulence due to spatial variations in $\alpha$ (possibly due to a
dead zone), may provide a means of preventing inward drift due to the
operation of corotation torques \citep{2001ApJ...558..453M,
2006ApJ...642..478M}.

\subsection{Migration of planetesimals - local model}

The local shearing box approximation that we have adopted in this work
does not include the effect of a radial pressure gradient in modifying
the orbital angular velocity of the gas. Consequently, the gas and
embedded bodies orbit at the same mean angular velocity, and
planetesimals do not undergo radial migration due to gas drag
interaction with a sub-Keplerian disc.  Large planetesimals, whose
interaction with the gas disc is gravity dominated, experience
stochastic gravitational forces that cause diffusion of the semimajor
axes (or equivalently, the guiding centres of the particle epicycles).
Smaller bodies, whose evolution is dominated by gas drag, similarly
diffuse radially due to the stochastic gas drag forces.  We discuss
both of these regimes below.

\subsubsection{Diffusion of gravity dominated bodies}

As with the larger planetesimals discussed in
Sect.~\ref{global-migration} for the global models, the larger
planetesimals in the shearing box simulations undergo radial diffusion
due to the stochastic gravitational forces they experience. The
deviation in the positions of the guiding centres of particles from
their initial values is denoted as $\Delta x$ (measured in units of
the local scale height $H$), and we plot the standard deviation of
this quantity for the gravity-only particles (set ``G") in
Fig.~\ref{fig:vdist_dx}.  As with the global simulations, the
evolution of $\Delta x$ displays random walk behaviour.  As is
illustrated by the fitted curve and error bounds in
Fig.~\ref{fig:vdist_dx}, the random walk can reasonably well
approximated by the function
\begin{equation}
  \sigma(\Delta x) \approx C_\sigma(\Delta x)\,\sqrt{t-t_0}\,
  \label{eqn:rwalk-fit}
\end{equation}
with a single fitting constant $C_\sigma=7.67 \pm 1.28 \times 10^{-3}$ 
representing the strength of the stirring mechanism.

\begin{figure}
  \center\includegraphics[width=\columnwidth]{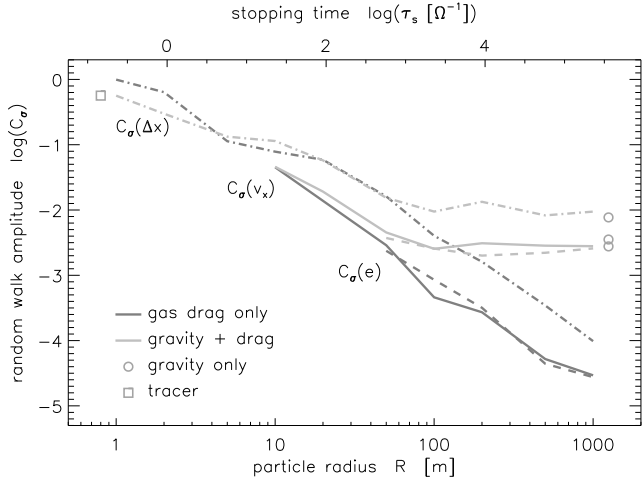}
  \caption{Excitation amplitudes for the random walk behaviour shown
    in Fig.~\ref{fig:vdist_dx}; lines correspond to $C_\sigma$ for the
    displacement (\emph{dash-dotted}), the rms velocity of the guiding
    centres (\emph{solid}), and the eccentricity (\emph{dashed
    lines}). Units are in $[H]$, $[\Cs]$, and $[H/R]$ times
    $[(2\pi/\Omega])^{\nicefrac{-1}{2}}]$, respectively.}
  \label{fig:rwalk}
\end{figure}

The amplitude of this fitting constant for the G+D particle set is
plotted using the light grey dash-dotted curve in Fig.~\ref{fig:rwalk}
(which also displays random-walk fitting parameters for the growth of
the radial velocity dispersion and eccentricity.  These values are
also listed in table~\ref{tab:rwalk}).  It is clear that for particles
with $\Rp \ge 100\m$, this coefficient has an almost constant value,
indicating that the radial diffusion of these larger sized
planetesimals is dominated by gravitational interaction with the disc.
The transition region is rather narrow, and G+D particles quickly
approach the values corresponding to the gravity only particle set.

\begin{table*}
  \begin{tabular}{llcccccccc}
    \hline
    & & gravity only & \multicolumn{3}{c}{gravity + drag} & 
    \multicolumn{3}{c}{drag force only} & tracer \\
    \hfill$[(2\pi/\Omega)^{-\nicefrac{1}{2}}]\,\times$ & 
      & & $10\m$ & $100\m$ & $1\km$ &$10\m$ & $100\m$ & $1\km$ & \\
    \hline
    $C_\sigma(\Delta x)$\hfill$\quad[10^{-2}\,H]$ 
    & & 0.77$\pm$0.13 & 11.4$\pm$0.02 & 0.95$\pm$0.11 & 0.94$\pm$0.10 & 
        7.71$\pm$0.03 & 0.44$\pm$0.12 & 0.01$\pm$0.18 & 56.5$\pm$0.01 \\
    $C_\sigma(v^{\rm gc}_x)$\hfill$\quad[10^{-3}\,\Cs]$ 
    & & 3.53$\pm$0.03 & 45.9$\pm$3.75 & 2.54$\pm$0.36 & 2.79$\pm$0.04 &
        45.1$\pm$3.10 & 0.46$\pm$0.01 & 0.03$\pm$0.00 & --- \\
    $C_\sigma(e)$\hfill$\quad[10^{-3}\,H/R]$
    & & 2.77$\pm$0.53 & --- & 2.57$\pm$0.04 & 2.58$\pm$0.01 &
                        --- & 0.85$\pm$0.01 & 0.03$\pm$0.00 & --- \\
    \hline
  \end{tabular}
  \caption{Random walk-amplitudes $C_\sigma$ describing the stochastic
    excitation (as depicted in Fig.~\ref{fig:vdist_dx}) of the
    dispersion within an ensemble. Values are obtained for the
    displacement $\Delta x$, the rms radial velocity dispersion,
    $v_x$, and the eccentricity $e$. With the exception of the first
    column, the applicability of the functional form $C_\sigma
    (t-t_0)^{1/2}$ is of course limited to the time interval before
    saturation is reached (see Figs.~\ref{fig:vdist_cmp1} and
    \ref{fig:vdist_cmp2}).}
  \label{tab:rwalk}
\end{table*}

We now compare the results shown in Fig.~\ref{fig:vdist_dx} with those
obtained for the global simulations.  The deviation of the position of
the particle guiding centres is measured in units of the local scale
height, $H$.  We see from Fig.~\ref{fig:vdist_dx} that the best fit
reaches a value $\Delta x = 0.107 H$ after a time $t=200$ orbits, and
would reach a value $\Delta x = 0.262 H$ after 1200 orbits.  Dividing
by the initial radial location of the particles, we obtain $\Delta x/r
= 0.262 H/r$. The shearing box runs have a value $H/r=0.075$, when we
consider that they are located at an orbital radius $r=5\au$, such
that $\Delta x/r =0.02$. This compares reasonably well with the
results of the global simulations, where we obtain a value of $\Delta
a/a = 0.03$ for model G1 at $t=1200$ orbits.  In terms of physical
parameters, model G5 is the most similar to the shearing box
simulation, but as may be seen in Fig.~\ref{migration-G1-and-G4}, the
rate of radial diffusion for models G1 and G5 are very similar.

As discussed in Sect.~\ref{global-migration}, the diffusion
coefficient associated with the random walk in radius of the particles
(or equivalently the diffusion of angular momentum) $D_j = \sigma_T^2
\tau_{\rm corr}$, where $\sigma_T$ is the standard deviation of the
fluctuating torque and $\tau_{\rm corr}$ is the torque correlation
time, averaged over the ensemble.  The diffusion coefficient found in
the global simulations was found to be in reasonable agreement with
the degree of particle diffusion obtained in those models.  For the
shearing box model, the value of $\sigma_T$ is given in
Fig.~\ref{fig:trq_dist}, and is found to be almost identical to that
obtained in the global model G1.  The torque correlation time is given
by Fig.~\ref{fig:ACFs}, and has a value $\tau_{\rm corr}=0.32$ orbits.
We thus see that the value of $\Delta x/r =0.02$ obtained by
extrapolating Fig.~\ref{fig:vdist_dx} to a time $t=1200$ orbits is in
very good agreement with what would be expected from the diffusion
coefficient $D_j = \sigma_T^2 \tau_{\rm corr}$, the prediction being
$\Delta x/r =0.021$.

\subsubsection{Diffusion of gas drag dominated bodies}
\label{sec:schmidt}

As with the gravity dominated particles, the gas drag dominated
particles diffuse radially over time. The radial displacement, $\Delta
x$ evolves according to a random walk, and can be reasonably well fit
by Eq.(~\ref{eqn:rwalk-fit}).  The fitting coefficients,
$C_\sigma(\Delta x)$, are plotted in Fig.~\ref{fig:rwalk} as a
function of planetesimal size.  In table~\ref{tab:rwalk} we
furthermore compile a collection of values for reference. We observe
that the diffusion via drag forces roughly scales with stopping time
like $\tau_{\rm s}^{-1}$ for $\tau_{\rm s}\ll 1$. When the stopping
time approaches the dynamical time $\Omega^{-1}$, which is the case
for boulders of a few metres in radius, the dependence on $\tau_{\rm
s}$ becomes weaker because the particles are tightly coupled to the
turbulent motion of the flow, and essentially behave like massless
tracers.  In this limit we see that, and $C_\sigma(\Delta x)$
approaches the value of the tracer particles in Fig.~\ref{fig:rwalk}.


To interpret the above findings in a more general context and be able
to scale the obtained values with respect to model parameters, we now
consider the Schmidt number, denoted $\Sc=\nu/D$.\footnote{To avoid
further confusion, we adhere to the conventions introduced in
Sect. 2.1 of \cite{2007Icar..192..588Y}.}  This is the ratio of the
momentum diffusion rate and the mass diffusion rate, and in our case
the momentum diffusion $\nu$ is given by the turbulent Maxwell and
Reynolds stresses as quantified by the dimensionless
$\alpha$~parameter. For the mass diffusion, we distinguish between
Lagrangian fluid elements (represented by massless tracer particles)
for which we obtain a diffusion coefficient $D_{\rm g}$, and
planetesimals with inertia, to which we assign the notation $D_{\rm
p}$.


\paragraph*{Lagrangian tracers:}

Before looking at the diffusion of the planetesimals themselves, we
need to provide a reference value. This is given by $\Sc_{\rm
g}=\nu/D_{\rm g}$, i.e., the Schmidt number for Lagrangian fluid
elements. To determine the diffusion coefficient $D_{\rm g}$, we
closely follow the approach described in Sect.~4 of
\cite{2006A&A...452..751F}. This means, we compute the velocity
autocorrelation function
\begin{equation}
  S^{ij}(\tau) = \rms{v_i(z(z_0,\tau),\tau)\,v_j(z_0,0)}\,,\label{eq:acf}
\end{equation}
where the dependence of $v_i$ on $z(z_0,\tau)$ implies that we are
considering the correlation with respect to a Lagrangian fluid
element. Note that \citeauthor{2006A&A...452..751F} have approximated
this by $v_i(z_0,\tau)$, i.e., the Eulerian velocity components
(cf. their Sect.~2.3). Because we include the evolution of massless
tracer particles, we do not rely on this approximation and can
directly compute Eq.(\ref{eq:acf}) along the particle trajectories.

\begin{figure}
  \begin{center}
    \includegraphics[height=0.52\columnwidth]{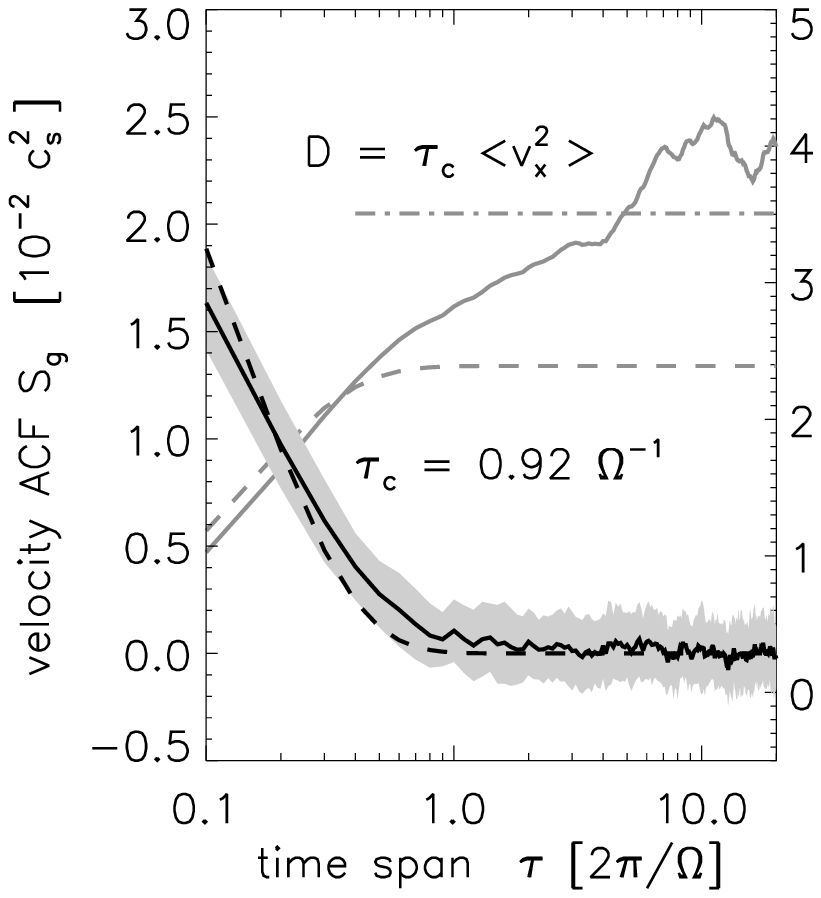}%
    \hfill
    \includegraphics[height=0.52\columnwidth]{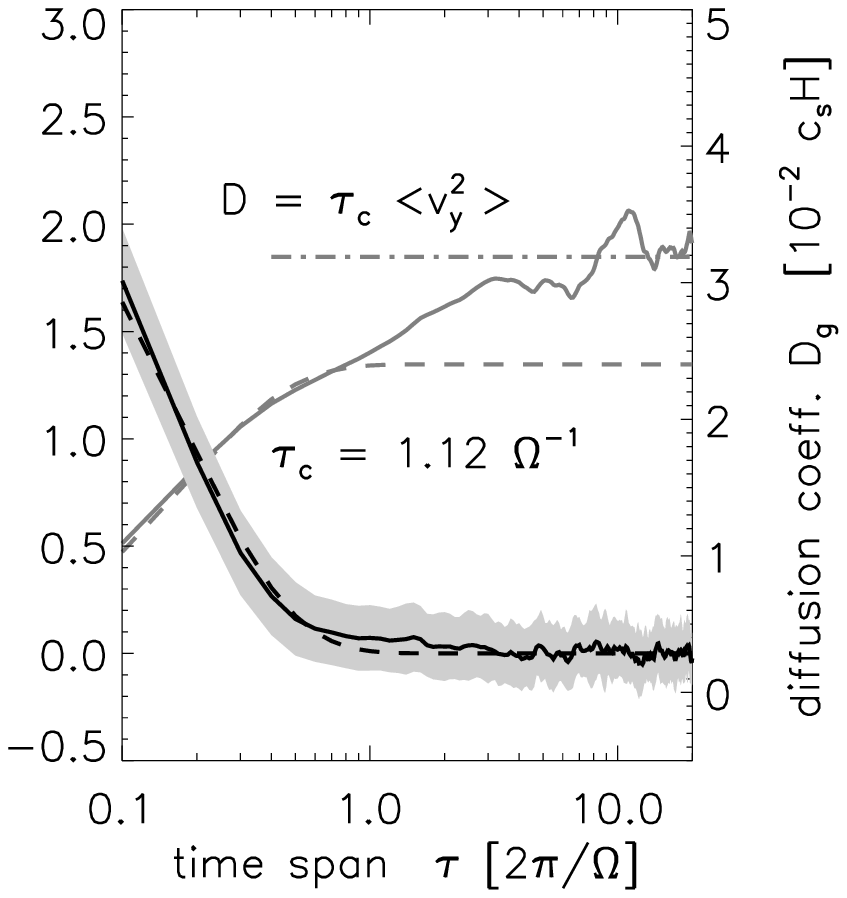}
  \end{center}
  \caption{Velocity ACFs $S_{\rm g}^{xx}$ (\emph{left panel}), and
    $S_{\rm g}^{yy}$ (\emph{right panel}) for massless tracers. From
    the ACFs (\emph{black lines, lhs axes}), we obtain time-integrated
    diffusion coefficients $D_{\rm g}(t)$ (\emph{grey line, rhs
    axes}). The fits $S_{\!\rm g}\sim\,{\rm e}^{-\tau/\tau_{\rm c}}$
    (\emph{dashed black}) result in corresponding diffusion profiles
    (\emph{dashed grey}) which agree for $\tau\simlt 2\pi/\Omega$. For
    $\tau\gg 1$, $D_{\rm g}$ can be approximated by the given
    theoretical estimate (\emph{dash-dotted line}).}
    \label{fig:sc_hydro} 
\end{figure}

The diffusion of a fluid element along the radial direction can then be
obtained by time-integrating the corresponding component of the
correlation tensor:
\begin{equation}
  D_{\rm g}(\tau) = \int_{0}^{\tau}{S^{xx}_{\rm \!g}(\tau')\,{\rm d}\tau'}\,.
  \label{eq:diff}
\end{equation}
As discussed in \cite{2006A&A...452..751F}, for large $\tau$ this can
be approximated by the product of the square of the rms velocity
$\rms{v_{\!x}^2} = S^{xx}_{\rm \!g}(0)$ and the correlation time
$\tau_{\rm c}$ of the turbulence. This estimate is based on the
assumption that the correlations in velocity decay like ${\rm
e}^{-\tau/\tau_{\rm c}}$. To check whether this assumption is
warranted, we plot $S_{\rm \!g}(\tau)$ and the derived $D_{\rm
g}(\tau)$. This is done separately for the $x$ and $y$ directions in
the left and right panel of Fig.~\ref{fig:sc_hydro}.

Looking at the two ``hockey stick'' shaped curves, we see that the
${\rm e}^{-\tau/\tau_{\rm c}}$ law (dashed lines) is met well for the
azimuthal component (right panel), whereas there is considerable
tension for the radial direction (left panel). Note that the time
domain has logarithmic scaling, which implies that the deviation in
the shape of the curve cannot be accounted for by changing the fit
parameter $\tau_{\rm c}$. Tentatively, an exponential decay law can be
restored via transformation to a generalised time variable
$\tilde{\tau}\rightarrow\tau^\alpha$ with $\alpha\simeq 0.7$ -- we
however leave it open whether the observed issue bears any relevance
at the current stage of modelling.

Our fitted correlation times are $\tau_{\rm c}^{(x)}= 0.92\,
\Omega^{-1}= 0.15\tms 2\pi\Omega^{-1}$ and $\tau_{\rm c}^{(y)}=1.12\,
\Omega^{-1}= 0.18\tms 2\pi\Omega^{-1}$. The obtained numbers are in
fact very close to the value of $0.15$ orbits, reported in
\cite{2006A&A...452..751F}. This supports the notion that $\tau_{\rm
  c}$ is characteristic for MRI-typical Mach numbers and only weakly
depends on the actual amplitude of the turbulent stresses.

The relation from Eq.~(\ref{eq:diff}) is illustrated by solid grey
lines in Fig.~\ref{fig:sc_hydro}, which represent the diffusion
``constant'' $D_{\rm g}$ as a function of time. For comparison, we
also plot the diffusion profile corresponding to the fitted solution
(dashed grey lines). These curves reasonably approximate the $\tau$
dependence of $D_{\rm g}$ for $\tau\simlt 2\pi/\Omega$. Note, however,
that the fitted amplitude is somewhat smaller than the rms velocities
and the curves saturate at a lower value. This means that the real
correlation functions have a (stochastic) tail that leads to a further
growth in $D_{\rm g}$ for $\tau\simgt 2\pi/\Omega$.

For times sufficiently greater than the correlation time of the flow,
the diffusion coefficient approaches a constant value approximated by
$\tau_{\rm c}\,\rms{v^2} \simeq 0.035$ as indicated by dashed-dotted
lines in the two panels of Fig.~\ref{fig:sc_hydro}. In accordance with
a recent study by \citet{2009arXiv0906.3314M}, the observed anisotropy
in the stream-wise and cross-stream directions is rather weak.

Translated into a Schmidt number, we yield a value of $\Sc_{\rm
g}\simeq 1.6$, which is considerably lower than the ratio between the
Maxwell and Reynolds stress, which fluctuates between $3$ and $4$. Our
value is a factor of about two smaller than the Eulerian value for
\emph{vertical} diffusion found by \cite{2006A&A...452..751F}. Schmidt
numbers obtained for small particles are usually inferred to be of
order unity \citep{2005ApJ...634.1353J,2006ApJ...639.1218T}, and our
results are compatible with this. 

A Schmidt number of order unity implies that small dust grains that
are strongly coupled to the gas will diffuse over large distances
during protostellar disc life times. Observations of crystalline
silicates embedded in circumstellar discs at significant distances
from their host stars, where temperatures are too low to explain
in-situ crystallisation of amorphous grains
\citep{2004Natur.432..479V}, suggest that turbulent diffusion may
indeed be responsible for transporting grains from the hot inner
regions of discs to the cooler outer regions. It remains to be
demonstrated, however, in a global, turbulent disc model, whether or
not a substantial number of grains can be transported outward against
the net inward mass accretion flow onto the star.


\paragraph*{Particles with inertia:}

Unlike massless tracers, real particles are subject to the additional
body forces in the rotating coordinate frame. Consequently, this leads
to the excitation of epicyclic oscillations. In the case of Keplerian
rotation, the associated angular frequency $\kappa$ is identical with
the local rotation rate $\Omega$. The autocorrelation function for a
perfect epicyclic oscillation is then simply proportional to
$\cos(2\pi\tau\Omega^{-1})$. This means that velocities are maximally
correlated if they are apart by a full period, and anti-correlated in
between.

\begin{figure}
  \center\includegraphics[width=0.9\columnwidth]{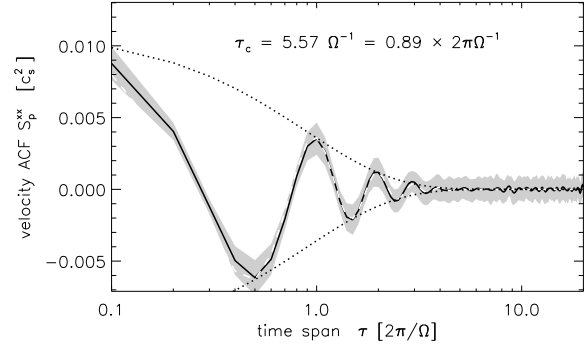}
  \caption{Velocity autocorrelation function $S_{\rm g}(\tau)$
    (\emph{solid black line}) for particles with radius $R=5\m$. The
    exponential decay is now modulated by epicyclic oscillations and
    can be fitted via a function $\sim\cos(2\pi\tau\Omega^{-1})\,{\rm
    e}^{-\tau/\tau_{\rm c}}$ (\emph{white dashed line}). The envelope
    (\emph{dotted lines}) is assumed to be relevant for the diffusion
    of the guiding centre.}
  \label{fig:sc_part}
\end{figure}

In general, the particles' motion will however deviate from a perfect
epicycle -- or, in other words, the defining elements of the motion
will change stochastically. With both the eccentricity and the
position of the guiding centre fluctuating, the coherence of the
motion is lost for larger and larger times. This is illustrated in
Fig.~\ref{fig:sc_part}, where we plot the velocity ACF for particles
of size $\Rp=5\m$. The epicyclic nature of the particle motion is
clearly seen in the sinusoidal modulation of $S_{\rm \!p}$. If
integrated over time, the ``diffusion'' process would rather resemble
a shaking motion than a random walk.

As a result, this renders the direct integration approach illustrated
in Fig.~\ref{fig:sc_hydro} difficult. Taking the excellent agreement
with the functional dependence $\sim\cos(2\pi\tau\Omega^{-1})\,{\rm
e}^{-\tau/ \tau_{\rm c}}$, we however conjecture that the diffusive
part of the overall motion can be recovered from the exponential
envelope function (dotted line in Fig.~\ref{fig:sc_part}). In
Tab.~\ref{tab:schmidt}, we report this correlation time as a function
of particle radius $\Rp$ and compare\footnote{The ratio $\tau_{\rm
s}/\tau_{\rm c}$ should not be confused with the Stokes number
$\St=\tau_{\rm s}/\tau_{\rm e}$ which measures the particle coupling
with respect to the eddy velocity.} it to the stopping time $\tau_{\rm
s}$. The ratio $\tau_{\rm c}/\tau_{\rm s}$ is about unity for
$\Rp=5\m$ and roughly scales with the square-root of the radius.

\begin{table}
  \begin{tabular}{lrrrrr} \hline
    $\Rp\ =$                         & 1m & 2m & 5m & 10m & 20m \\\hline
    $\tau_{\rm s}\ [\Omega^{-1}]$  & 0.23 & 0.91 & 5.71 & 22.8 & 91.4 \\
    $\tau_{\rm c}\ [\Omega^{-1}]$  & 0.68 & 1.59 & 5.74 & 16.7 & 46.4 \\
    $\tau_{\rm s}/\tau_{\rm c}$    & 2.96 & 1.74 & 1.01 & 0.73 & 0.51 \\
    $\Sc_{\rm p}$                  & 4.22 & 27.5 & 45.3 & 96.4 & 182. \\
    \hline
  \end{tabular}
  \caption{Time scales and Schmidt numbers $\Sc_{\rm p}=\alpha_{\rm
    ss}\,\Cs H/D_{\rm p}$ for the diffusion of the guiding centres of
    heavy particles on excited epicyclic orbits.
    \label{tab:schmidt}}
\end{table}

\begin{figure}
  \center\includegraphics[width=0.9\columnwidth]{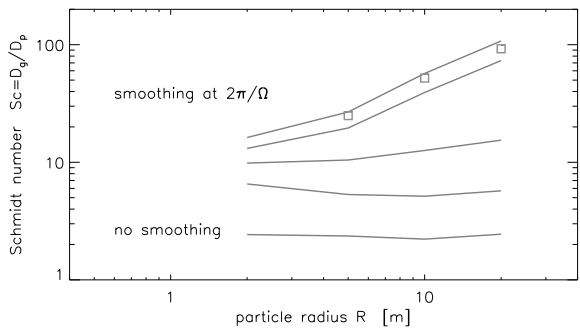}
  \caption{Attempt to determine the Schmidt number $\Sc=D_{\rm
      g}/D_{\rm p}$ for the diffusion of the guiding centres. The
      diffusion coefficient is estimated via $D_{\rm p}\simeq
      \tau_{\rm c} \rms{v_{\rm gc}^2}$, where $v_{\rm gc}$ is obtained
      by smoothing over the epicycles (\emph{solid lines}). An
      alternative estimate (\emph{open squares}) is derived
      independently via the coefficients $C^2_\sigma(\Delta x)$ in
      Tab.~\ref{tab:rwalk} -- note that these do not depend on any
      form of smoothing.}
  \label{fig:sc_epi}
\end{figure}

Given we can obtain a good estimate for the rms velocity of the
guiding centres, we can then compute $D_{\rm p} \simeq \tau_{\rm c}
\rms{v_{\rm gc}^2}$. Looking at Fig.~\ref{fig:xdisp}, it seems less
than straightforward to deconvolve the random motion of the guiding
centre from the overall motion, especially since the amplitude of the
epicycles changes on comparable time scales. As a first attempt, we
simply smooth-out the oscillations applying a box-car filter in
time. The thus derived Schmidt number $\Sc=D_{\rm g}/D_{\rm p}$ as a
function of particle size is plotted in Fig.~\ref{fig:sc_epi} for
various values of the filter size.

If we omit the smoothing, the results (although not strictly correct)
reflect the kinetic energy in the epicyclic mode itself. In this case,
the gradual decoupling of the particles implies that $\rms{v^2}$
decreases while $\tau_{\rm c}$ increases accordingly. Interestingly,
this leaves the overall particle diffusion rate unchanged. If we
increase the filter size, the epicyclic motion is attenuated and the
size dependence gradually becomes steeper. Reasonable convergence is
obtained when the filter size approaches the critical value, equal to
the epicyclic period. In Tab.~\ref{tab:schmidt} we list the
corresponding numbers.

Note that already for metre-sized objects the particle diffusivity is
reduced relative to Lagrangian tracers. We attribute this to the
effect that part of the kinetic energy is absorbed in ordered
epicyclic motion and is thus not available for particle diffusion
\citep[cf. Sect.~3.2 in][]{2007Icar..192..588Y}. The partial slippage
of the particles through the gas as a result of weaker drag
acceleration also causes the particle guiding centre velocities to be
reduced relative to the fluid turbulent velocities.

It might be seen as a deficiency of the approach that it depends on
the proper choice of a box-car filter function. To reinforce the validity
of the derived Schmidt numbers, we therefore provide an independent
estimate from the squares of the coefficients $C_\sigma(\Delta x)$
from Fig.~\ref{fig:rwalk}. Defining the growing spread in the
particles position with time, these can similarly be interpreted as a
diffusion constant. For consistency, we have also checked that the fit
constants $C_\sigma(\Delta x)$ do not change significantly, when the
epicyclic part of the motion is smoothed. Considering that this second
estimate solely relies on positions while the original one is derived
from velocity correlations, the agreement seen in
Fig.~\ref{fig:sc_epi} (where we plot the different estimates for the
Schmidt number) is quite striking.


\section{Conclusions}
\label{sec:Conclusions}

We have presented results from both local and global MHD simulations
which examine the evolution of planetesimals and boulders of different
size embedded in turbulent protoplanetary disc models.  The main aims
of this work are to: determine the magnitude of the velocity
dispersion which is excited in a planetesimal swarm; determine the
rate of radial diffusion of the planetesimals; determine under which
conditions broad agreement is obtained in the results from local
shearing box simulations and global disc simulations.

We find that the magnitude of density fluctuations, and the associated
stochastic forces experienced by embedded planetesimals is sensitive
to the dimensions of shearing box models. This appears to be related
to the fact that accurate modelling of the excitation and non-linear
steepening of spiral density waves requires boxes which are elongated
in the azimuthal direction \citep{2009MNRAS.397...64H}.  The two-point
correlation function for the perturbed surface density also shows that
coherent structures stretched by a distance $\simeq 6H$ are a common
feature of MHD turbulence in discs, indicating that shearing boxes in
excess of this length scale are required to prevent the premature
truncation of the gravitational field which results from these
structures.  In addition to the magnitude of the stochastic torques
being dependent on the box size, we also found that the correlation
time of the fluctuating torques also depends on the box size and
aspect ratio, with smaller boxes generating shorter correlation
times. This is apparently due to the ability of waves to propagate
across the box on multiple occasions due to the periodic boundary
conditions employed, combined with the rate at which they shear
past the planet due to the background flow. 
Shearing boxes with dimensions $4H \times 16H
\times 2H$ were found to provide converged results which agree well
with the results from global simulations.

We find that both global simulations and local shearing box
simulations predict that rapid excitation of planetesimal random
velocities is expected in fully turbulent disc models whose local
surface densities are similar to the minimum mass solar nebula. A
model whose turbulent stresses generate $\alpha =0.035$ leads to rapid
growth of the radial velocity dispersion, $\sigma(v_r)$, {\it via} a
random walk, such that after 1200 orbits at $5\au$
$\sigma(v_r)=200\ms$. A model with weaker turbulence ($\alpha =0.017$)
gave rise to $\sigma(v_r)=166\ms$ over the same evolution time.  These
random velocities are much larger than either the escape velocities
from planetesimals with sizes $1\km$ or $10\km$, or the catastrophic
disruption thresholds for collisions between bodies of similar size
\citep{1999Icar..142....5B, 2009ApJ...691L.133S}, suggesting that
planetesimal collisions occurring in fully turbulent discs similar to
those considered in this paper will result in destruction of the
planetesimals. We find that the expected equilibrium velocity
dispersion for $10\km$ sized planetesimals scales weakly with the
turbulent stress parameter such that $\sigma(v_r) = K_{v_r}
\alpha^{0.2}$. Extrapolation of the simulation results presented in
this paper to low values of $\alpha$ suggest that even with
$\alpha=10^{-6}$, bodies with size $10\km$ are likely to have random
velocities which will cause catastrophic disruption during mutual
collisions.  Values of $\sigma(v_r)$ small enough to allow runaway
growth of planetesimals to occur will be even more difficult to
achieve.  For smaller bodies, which are more tightly coupled to the
gas {\it via} gas drag, we find that the equilibrium velocity
dispersion decreases, and reaches a minimum of $\sigma(v_r) \simeq
20\ms$ for bodies of size $50\m$ in a disc with $\alpha=0.035$. For
smaller bodies than these the increasing influence of gas drag causes
the velocity dispersion to increase with decreasing size, and boulders
with $\Rp=1\m$ develop random velocities which closely match the
turbulent gas motions.

In addition to driving the growth of the velocity dispersion, or
equivalently the eccentricity, the stochastic forces experienced by
the planetesimals cause them to diffuse radially through the disc. For
large planetesimals with $\Rp=10\km$, radial drift through the disc
due to gas drag occurs very slowly, and their radial motion is
expected to be dominated by diffusion caused by the stochastic
gravitational field of the turbulent disc. Indeed, after a putative
disc life time of $5\Myr$, the rms relative change in semimajor axis,
$\sigma(\Delta a/a)$ for a planetesimal swarm located at $5\au$ in a
disc with $\alpha=0.035$ will be 50\%, which would have been
sufficient to cause large scale migration of the small body
populations of the Solar System during its formation. The fact that
such large scale migration of large planetesimals does not appear to
have happened allows us to put constraints on the strength of $\alpha$
in the solar nebula. A value of $\alpha \simeq 10^{-5}$ would lead to a more
reasonable $\sigma(\Delta a/a) \simeq 0.1$, consistent with the notion
that the solar nebula had a dead zone in the region where planet
formation took place.

Smaller planetesimals with $\Rp=10$ - $100\m$ are expected to undergo
large-scale radial drift due to gas drag, and this is observed in our
simulations. The migration rate of $10\m$ size bodies in the turbulent
discs, relative to that expected for an equivalent laminar disc model,
is found to be very similar. Migration in turbulent discs is found to
be slower by a factor of approximately 2 - 3, and this is due to
radial structuring of the disc's density profile, creating regions
where the pressure gradient (and hence rotational velocity profile) is
modified.

There have been a number of previous studies of the dynamical
evolution of planetary bodies embedded in turbulent
discs. \citet{2004MNRAS.350..849N} examined the torques experienced by
low mass planets and suggested that random walk behaviour for such
bodies should be expected, as has been observed in this
paper. \citet{2005A&A...443.1067N} examined the orbital evolution of
planets embedded in turbulent discs, with particular emphasis on the
semimajor axis and eccentricity changes. The strength of the
turbulence in that model was somewhat weaker than considered in this
paper ($\alpha \simeq 5 \times 10^{-3}$), although the disc model was
somewhat more massive, and in general it was found that the response
of embedded bodies to the stochastic forcing was stronger in that
study, with larger changes in semimajor axis and eccentricity being
observed. The primary reason for this discrepancy appears to be the
fact that a persistent vortex formed in the disc model considered by
\citet{2005A&A...443.1067N}.  Such vortices were also reported in the
work of \citet{2005MNRAS.364L..81F}, and interaction between embedded
bodies and such structures can lead to significant modification of
the semimajor axes and excitation of eccentricity. The models
presented in this paper do not have such vortex features, but during
the early stages of this project disc models with lower values of
$\alpha$ were found to generate such flow features quite readily.  One
possibility is that global disc models which are close to the limits
of resolving the MRI generate substantial variations in $\alpha$ as a
function of radius, which in turn generate localised pressure maxima
which are particularly prone to the formation of vortices
\citep{1987MNRAS.225..677H, 1999ApJ...513..805L}.  
Models with smaller values of the plasma
$\beta$ parameter have stronger field strengths and thus resolve the
MRI more easily, and in addition the stronger fields may act to
inhibit vortex formation through the action of local magnetic
stresses.  It is these models that we have considered in this paper.

\citet{2009ApJ...707.1233Y} recently considered the evolution of
planetesimal swarms using non stratified shearing box simulations. The
planetesimals in that study experienced the fluctuating gravitational
field of the disc, but did not experience gas drag forces. Although
random walk behaviour of the particles in that study was observed, in
general it was found to be significantly weaker than we report here,
and one conclusion reached by these authors is that turbulence is
unlikely to cause the catastrophic disruption of planetesimals.  The
primary reason for this appears to be the fact that most simulations
performed by \citet{2009ApJ...707.1233Y} used shearing boxes with $2H
\times 2H \times 2H$.  Test calculations with larger boxes presented
in that paper indicated a strong increase in the response of the
particles to the turbulence forcing as a function of increasing box
size, suggesting that the main reason for the discrepancy in our
results is due to this effect.

\citet{2008ApJ...686.1292I} recently considered the evolution of
planetesimals embedded in turbulent protoplanetary discs by means of
N-body simulations combined with a prescription for planetesimal
stirring based on the work of \citet{2004ApJ...608..489L}. In basic
agreement with the results we have presented in this paper, they show
that turbulence leads to the excitation of a large velocity
dispersion, which is likely to cause catastrophic disruption of
planetesimals rather than growth following mutual collisions, for a
wide range of turbulent strengths.

The simulations we have presented here use the simplest possible
numerical set-up: ideal MHD in non stratified disc models. In a future
paper we will present the results of a similar study using vertically
stratified models with and without dead zones. This future study will
tell us whether we need to examine new paradigms for the rapid growth
of planetesimals, or whether instead runaway growth can indeed occur
within a dead zone of an otherwise turbulent protoplanetary disc.

\section*{Acknowledgements}
O.G. thanks Sebastien Fromang for discussions on improving the
numerical scheme and providing his spectral analysis tools. 
R.P.N thanks Alessandro Morbidelli and Bill Bottke for 
discussions about the radial mixing of asteroids. This work
used the \NIII code developed by Udo Ziegler at the Astrophysical
Institute Potsdam. All computations were performed on the QMUL HPC
facility, purchased under the SRIF initiative. R.P.N. and O.G.
acknowledge the hospitality of the Isaac Newton Institute for
Mathematical Sciences, where most of the work presented in this paper
was completed during the `Dynamics of Discs and Planets' research
programme.

\appendix
\bsp

\label{lastpage}

\end{document}